\documentclass[12pt,preprint]{aastex}
\usepackage{subfigure}
\begin{document}

\title{{\it Spitzer} Spectroscopy of Circumstellar Disks in the 5\,Myr Old Upper Scorpius OB Association}

\author{S. E. Dahm\altaffilmark{1} \& John M. Carpenter\altaffilmark{2}}

\altaffiltext{1}{W. M. Keck Observatory, 65-1120 Mamalahoa Hwy, Kamuela, HI 96743}
\altaffiltext{2}{Department of Astronomy, California Institute of Technology, MS 105-24, Pasadena, CA 91125}

\begin{abstract}
We present mid-infrared spectra between 5.2 and 38 $\mu$m for 26 disk-bearing members of the $\sim$5\,Myr 
old Upper Scorpius OB association obtained with the Infrared Spectrograph (IRS) onboard the {\it Spitzer}
Space Telescope. We find clear evidence for changes in the spectral characteristics of dust emission
between the early (B+A) and late-type (K+M) infrared excess stars. The early-type members exhibit featureless 
continuum excesses that become apparent redward of $\sim$8 $\mu$m. In contrast, 10 and 20 $\mu$m silicate
features or PAH emission are present in all but one of the late-type excess members of Upper Scorpius.
The strength of silicate emission among late-type Upper Scorpius members is spectral type dependent, 
with the most prominent features being associated with K5--M2 type stars. By fitting the spectral energy
distributions (SED) of a representative sample of low-mass stars with accretion disk models, we find that 
the SEDs are consistent with models having inner disk radii ranging from $\sim$0.2 to 1.2 AU. Complementary high
resolution ($R\sim33,000$) optical ($\lambda$$\lambda$4800--9200) spectra for the Upper Scorpius excess stars 
were examined for signatures of gaseous accretion. Of the 35 infrared excess stars identified in Upper Scorpius,
only 7 (all late-type) exhibit definitive signatures of accretion. Mass accretion rates ($\dot{M}$) for
these stars were estimated to range from $10^{-11}$ to $10^{-8.9}$ M$_{\odot}$ yr$^{-1}$. Compared to Class II
sources in Taurus-Auriga, the disk population in Upper Scorpius exhibits reduced levels of near and 
mid-infrared excess emission and an order of magnitude lower mass accretion rates. These results suggest
that the disk structure has changed significantly over the 2--4\,Myr in age separating these two stellar
populations. The ubiquity of depleted inner disks in the Upper Scorpius excess sample implies that such
disks are a common evolutionary pathway that persists for some time.
\end{abstract}


\section{Introduction}

Following initial gravitational collapse, most protostars from $\sim$10 M$_{\odot}$ to the hydrogen burning limit 
are surrounded by optically thick disks of gas and dust, the progenitors of planetary systems. Early ground-based
infrared ($JHKLN-$band) observations established that the inner disks ($\ll$ 1.0 AU) around most stars 
have dispersed by an age of $\sim$10\,Myr (Haisch et al. 2001; Mamajek et al. 2004). Observations with the 
{\it Spitzer} Space Telescope (Werner et al. 2004) have also demonstrated that just a few percent of near
solar-mass stars older than $\sim$10\,Myr exhibit dust emission from the inner $\sim$3 AU of the disk
(Uchida et al. 2004; Silverstone et al. 2006; Hernandez et al. 2007). Over similar timescales, mass accretion
rates decline by at least an order of magnitude or are halted altogether (Muzerolle et al. 2000; Hartmann 2005). 
Growing evidence from {\it Spitzer} suggests that disk evolution is also strongly mass dependent
(Lada et al. 2006; Carpenter et al. 2006; Dahm \& Hillenbrand 2007), such that low-mass stars (i.e. solar-mass and less massive) 
are capable of retaining optically thick circumstellar disks for prolonged periods relative to their more massive counterparts.

The Upper Scorpius OB association possesses several critical attributes that make it an important region for studies
of disk evolution. Located $\sim$145 pc distant in the local spiral arm, Upper Scorpius is among the nearest OB associations to the Sun
(Blauuw 1991; de Zeeuw et al. 1999; Preibisch \& Zinnecker 1999; Preibisch et al. 2002). With an age of $\sim$5\,Myr,
it is the youngest of 3 sub-groups comprising the greater Scorpius OB2 or Scorpius-Centaurus OB association 
(de Zeeuw et al. 1999; Preibisch et al. 2002). Upper Centaurus-Lupus and Lower Centaurus-Crux, the other components 
of Scorpius OB2, have median ages of $\sim$17\,Myr (Mamajek et al. 2002). Early-type members (B--G spectral types) 
of Upper Scorpius were identified by de Zeeuw et al. (1999) in their {\it Hipparcos} proper motion survey of the 
association. A total of 120 stars (49 B, 34 A, 22 F, 9 G, 4 K, and 2 M-type) were confirmed as members based 
upon their measured proper motions. Walter et al. (1994) conducted the first survey for low-mass (G--M spectral 
types) members of Upper Scorpius by obtaining follow-up, high resolution spectroscopy and near-infrared imaging of 
69 {\it Einstein} X-ray sources spread over 7 square degrees. Of these, 28 were confirmed as pre-main sequence stars. 
Preibisch \& Zinnecker (1999) identified $\sim$100 low-mass members of Upper Scorpius from a wide-field (160 square 
degrees) ROSAT X-ray survey of the association. The pre-main sequence stars were found to have a median age of $\sim$5\,Myr
and exhibit no evidence for a significant age dispersion. An additional 68 pre-main sequence members including
many M-dwarfs were identified by Preibisch et al. (2002) using the 2dF multiobject spectrograph. Their findings
suggest that the mass function of Upper Scorpius is consistent with that of the Galactic disk population. In a
150 square degree photometric and spectroscopic survey of Upper Scorpius, Slesnick et al. (2008) identified
145 new low-mass members and placed an upper limit of $\pm$3 Myr on the age dispersion of the association
(see also Preibisch \& Zinnecker 1999).

In a 4--16 $\mu$m {\it Spitzer} survey of 204 members of the Upper Scorpius OB association with masses ranging 
from $\sim$0.1 to 20 M$_{\odot}$, Carpenter et al. (2006) find only 35 stars with emission in excess of photospheric 
values at 8 or 16 $\mu$m. The low-mass (0.1--1.2 M$_{\odot}$) stars are associated with excess emission that 
is characteristic of optically thick disks found around classical T Tauri stars. Stars 
more massive than $\sim$1.8 M$_{\odot}$ are found with weak excesses at short wavelengths indicating that the inner 
$\sim$10 AU are cleared of primordial dust. Such excesses are more analogous to second generation, optically-thin debris disks 
produced by the collision of planetesimals. Of particular interest, none of the $\sim$30 F or G-type stars (1.2--1.8 M$_{\odot}$) 
included in the {\it Spitzer} survey exhibit infrared excess emission at wavelengths $<$16 $\mu$m. These results
imply that the mechanisms responsible for disk dispersal operate more efficiently for high mass 
stars than for their low-mass counterparts. Consequently, longer timescales may be available for the buildup of 
planetesimals and planetary systems within the terrestrial zone for stars less massive than the Sun. The stars in 
Upper Scorpius also lack the short-wavelength ($JHK-$band) excess emission characteristic of younger T Tauri stars, 
implying that the inner disk regions close to the star are evacuated of dust grains. Prior to the launch 
of {\it Spitzer}, examples of sources with cleared inner disks (so called ``transition disks'') were relatively rare, 
but are now becoming increasingly common (e.g. CoKu Tau 4, GM Aur, DM Tau, TW Hya; see also Muzerolle et al. 2006).

Whether transition disks are a common, but short-lived evolutionary pathway or just one of many possible
intermediate stages of disk evolution and planet formation is unknown. Physical processes that have been proposed 
as disk clearing mechanisms include: a) dust grain growth and mid-plane settling which deplete the terrestrial 
region of micron-sized grains; b) binarity or the formation of a giant planet that dynamically clears the inner 
disk of gas and dust; and c) photoevaporation of the inner disk by photospheric and accretion-generated UV and FUV 
flux (Najita et al. 2007). A critical step toward understanding these transition disks is evaluating the spatial distribution of remnant 
gas. If grain growth is responsible for the depletion of small dust grains, a gaseous inner disk should remain and 
accretion is expected to continue unabated. If a massive planet has formed in the terrestrial region, gaseous accretion 
onto the stellar photosphere would either be significantly reduced or altogether terminated by the gravitational
influence of the planet, which prevents replenishment of inner disk gas. A photoevaporated disk will have similar
traits, but can be distinguished from dynamical clearing by overall disk mass (Alexander et al. 2006).

In order to probe dust disk structure and gaseous accretion, we have obtained {\it Spitzer} low resolution
mid-infrared spectra and Keck high resolution optical spectra for infrared excess sources in Upper Scorpius.
This $\sim$5\,Myr old disk-bearing population provides a sample that can be compared directly with 
the presumably younger Class II and III sources in the Taurus-Auriga star forming region. Identical data sets 
(photometry and IRS spectra) are available for both stellar populations, permitting a detailed examination 
of the effects of evolution upon disk structure. In \S2 we describe the Upper Scorpius
membership sample and provide details of the {\it Spitzer} IRS and the Keck High Resolution Echelle Spectrometer 
(HIRES) observations. We then discuss (\S3) the observed spectral energy distributions (SED) for the members observed 
with IRS. Next (\S4) we review accretion diagnostics and estimate $\dot{M}$ for the Upper Scorpius sources 
observed with HIRES. We then compare (\S5) the infrared excess properties of the Upper Scorpius disk sample
directly with the Taurus-Auriga Class II and Class III populations from Furlan et al. (2006). Finally, we 
examine (\S6) the dust properties of early-type Upper Scorpius excess members inferred from the photosphere-subtracted
SEDs. 

\section{Observations and Analysis}

\subsection{Upper Scorpius Membership Sample}
The membership sample of Carpenter et al. (2006) was compiled from {\it Hipparcos} astrometry for the early-type
stars (de Zeeuw et al. 1999), color-magnitude diagrams with follow-up \ion{Li}{1} $\lambda$6708 observations for
G--M type stars (Preibisch \& Zinnecker 1999; Preibisch et al. 2002), and X-ray detected G--M type stars with
\ion{Li}{1} $\lambda$6708 confirmation (Walter et al. 1994; Mart\'{i}n 1998; Preibisch et al. 1998; Kunkel 1999; 
K\"{o}hler et al. 2000). Given that the membership selection criteria are based upon stellar properties
(proper motions, X-ray activity, \ion{Li}{1} $\lambda$6708 absorption) unrelated to circumstellar disks,
it is believed that the sample is unbiased toward the presence or absence of disks. Carpenter et al. (2006)
empirically established thresholds to identify infrared excess sources, which were defined as exhibiting 8--4.5 $\mu$m
and 16--4.5 $\mu$m flux ratios that exceeded a fitted relation by 4 times the rms of the fit residuals and 4 times
the internal uncertainty in the specified flux ratio. Infrared excesses were detected for 35 sources; 29 at 8 $\mu$m 
and 33 at 16 $\mu$m. Their analysis finds that 24 of 127 (19\%) K and M-type stars exhibit 8 $\mu$m excesses,
5 of 61 (8\%) B and A-type exhibit similar excesses, while 0 of 30 F and G-type stars exhibit such excesses.
At 16 $\mu$m, the excess fraction is 23 of 121 (19\%) K and M-type stars, 10 of 52 (19\%) B and A-type stars,
and 0 of 22 F and G-type stars. When the IRS proposal was submitted, only 26 of the 35 excess sources were
known, given that not all IRAC and IRS 16 $\mu$m peak-up photometry were available. 

\subsection{{\it Spitzer} Infrared Spectrograph Observations}
{\it Spitzer} IRS low resolution spectra were obtained for 26 stars in Upper Scorpius that have an infrared
excess: 25 sources were observed for this study during campaign 1067 from 2006 September 14--19 and campaign
1085 from 2007 March 16--21, and one source ([PZ99]J161411.0-230536) was obtained from the FEPS {\it Spitzer}
Legacy program as described by Carpenter et al. (2008). The following discussion describes the data reduction
procedures for the 25 spectra obtained for this study.

The full 5.2--38 $\mu$m spectra were obtained using the two low-resolution ($\lambda$/$\delta$$\lambda$$\sim$60--120) 
IRS modules: Short-Low (SL) 5.2--14.5 $\mu$m and Long-Low (LL) 14.0--38.0 $\mu$m. All targets were
observed in IRS staring mode with ramp durations of 6 or 14 seconds. Multiple cycles
were performed to ensure detection of the stellar photosphere and for bad pixel rejection. The faintest 
mid-infrared source required 30 cycles in the SL module while the brightest only 2 cycles. All images were 
initially processed by the IRS data pipeline software, version S15.3.0. Rogue pixel were removed from the 
resulting basic calibrated data (bcd) images using the {\it irsclean} software package and the campaign based 
rogue pixel masks. The individual spectra were then extracted and calibrated using the Spectroscopic Modeling, 
Analysis and Reduction Tool (SMART; Higdon et al. 2004). Background sky subtraction was accomplished using 
observations made at the same nod position for each IRS module and order. The object spectra on the resulting 
sky-subtracted frames were then automatically identified, traced, and extracted using the tapered column point 
source extraction option that is recommended for low-resolution observations. In this extraction mode, the 
region of extracted pixels is tapered along each spectral order, and its width in the cross-dispersion 
direction is scaled with the instrumental point spread function. Given the single point source nature of the 
observed stars and the standard extraction aperture used, the SMART pipeline-produced flux calibration was adopted. 
Repeatability of the IRS flux calibration observation is stated to be at the 5\% level while absolute photometric accuracy is 
$\sim$10\%. Once extracted, the individual spectra were then cleaned of remaining aberrant pixels and coadded 
using tools available within the SMART IDEA main window. The coadded spectra were then imported into IDL and 
median combined to produce the final flux calibrated, merged spectra presented here.

\subsection{High Resolution Echelle Spectrometer}
The High Resolution Echelle Spectrometer (HIRES: Vogt et al. 1994) on Keck I was used on 2006 June 16 and 
2007 May 24--25 to obtain spectra of the 35 Upper Scorpius stars identified by Carpenter et al. (2006) that
have an 8 and/or 16 $\mu$m excess. Conditions were photometric 
for all 3 nights with seeing conditions varying from 0.6--0.9\arcsec. HIRES was used with the red cross-disperser
and the C5 decker (1.148\arcsec$\times$7.0\arcsec), which has a projected slit width of 4 pixels and a spectral
resolution of $\sim$33,000 (8.8 km s$^{-1}$). The cross-disperser and echelle angles were set to approximately
0.884$^{\circ}$ and 0.0$^{\circ}$, respectively, providing nearly complete spectral coverage from $\sim$4800--9200 \AA.
The selected wavelength range includes several gravity and temperature-sensitive photospheric features as well
as permitted and forbidden transitions generally associated with accretion processes or chromospheric activity:
H$\beta$, \ion{He}{1} $\lambda$5876, [O I] $\lambda$6300, H$\alpha$, [S II] $\lambda\lambda$6717, 6731, and
\ion{Ca}{2} $\lambda\lambda$8498, 8542, and 8662. The red, green, and blue detectors were used in low gain mode,
resulting in readout noise levels of 2.8, 3.1, and 3.1 e$^{-1}$, respectively. Internal quartz lamps were used
for flat fielding and ThAr lamp spectra were used for wavelength calibration. The HIRES data were reduced using
the MAuna Kea Echelle Extraction, {\it makee}, reduction script written by Tom Barlow. The spectra were not
flux calibrated. 
Given the range of apparent magnitudes for the observed sample, integration times varied from 15 to 45 minutes
with typical signal-to-noise ratios of $\sim$30, 50, and 100 being achieved on the blue, green, and red chips,
respectively. In Table 1, we list the measured spectroscopic properties of the Upper Scorpius membership sample
observed with HIRES. The columns tabulate the source name from Carpenter et al. (2006), assumed spectral type, adopted extinction value from
Preibisch \& Zinnecker (1999) or Preibisch et al. (2002), measured radial velocities with 1 $\sigma$ uncertainties,
and measured equivalent widths for H$\beta$, \ion{He}{1} $\lambda$5876, H$\alpha$, \ion{Li}{1} $\lambda$6708, and
\ion{Ca}{2} $\lambda$8542. Measurement uncertainties for equivalent widths are $\sim$0.05\AA. Other emission lines 
(e.g. [O I] $\lambda$6300) found within the observed spectra are also tabulated. 

\section{{\it Spitzer} Mid-Infrared Spectra of Upper Scorpius Members}

\subsection{Spectral Energy Distributions}
In Figures 1a--1d, we present the SEDs of the 35 Upper Scorpius sources with an infrared excess.
The mid-infrared spectra extending from $\sim$5.2 to 38 $\mu$m for the 26 stars observed with IRS 
are plotted in red and boxcar smoothed to reduce noise.
Superimposed on the SEDs are extinction-corrected optical ($BVRI$) and near-infrared ($JHK_{S}$) photometry from
the literature and 2MASS, and {\it Spitzer} IRAC 4.5 and 8.0 $\mu$m fluxes and IRS peak-up photometry
from Carpenter et al. (2006). Also shown are the MIPS 24 and 70 $\mu$m fluxes or 3$\sigma$ upper
limits from Carpenter et al. (2009, in preparation), which place strong constraints upon the shape of the SED beyond the
IRS wavelength range. Given the minimal interstellar reddening suffered by the Upper Scorpius
members ($E_{B-V}$ $\le$ 0.94 mag), only the optical and near-infrared photometry are corrected for extinction 
in the SEDs. Also shown in Figures 1a--1d are the NextGen stellar atmospheric models of Hauschildt et al. (1999)
for solar metallicities and sub-giant surface gravities (log g $= 4.0$). The main sequence effective temperatures 
($T_{\rm eff}$) from Kenyon \& Hartmann (1995) are adopted for the assigned spectral types of the Upper Scorpius 
members. The NextGen models are available over a range of $T_{\rm eff}$ values from 3,000 to 10,000 K at a 
resolution of 200 K, which is adequate coverage and granularity for the Upper Scorpius membership sample. The model 
stellar photospheres in Figures 1a--1d are normalized to the $J-$band flux for late-type stars (K+M) and to 
the $V-$band fluxes for the early-type (B+A) members, which is near the peak of the stellar photospheric emission.
To facilitate comparison of the spectra and the strengths of the 10 and 20 $\mu$m silicate emission features,
we present only the IRS data in Figures 2--4, grouped by spectral type. The figures clearly demonstrate the 
diversity of observed infrared excess properties of the Upper Scorpius disk sample.

The mid-infrared spectra of all disk candidates show definitive excesses indicative of circumstellar 
dust emission. The early-type (B+A) Upper Scorpius members (Figs. 1a, 1b, \& 2) exhibit near-photospheric SEDs out to
$\sim$8 $\mu$m, beyond which a weak excess becomes discernable. The polycyclic aromatic hydrocarbon (PAH)
emission and the 10 and 20 $\mu$m silicate emission features that often dominate the mid-infrared spectra of 
Herbig AeBe stars (Keller et al. 2008; Sloan et al. 2005) are absent in these early-type members. While
most early-type Upper Scorpius excess stars exhibit declining flux levels from 5.2 to 38 $\mu$m, HIP 76310
and HIP 80088 show flattened or slightly rising mid-infrared SEDs beyond $\sim$10 $\mu$m. 
These two stars were the only early-type excess members observed with IRS that were
detected by the MIPS 70 $\mu$m survey of Carpenter et al. (2009, in preparation). HIP 80024, which also appears to exhibit a
slightly rising mid-infrared SED, was not detected at 70 $\mu$m by MIPS. 

The late-type (K+M) excess members of Upper Scorpius present strikingly different SEDs in Figures 1b, 1c, and 1d 
than their more massive counterparts. Nearly all exhibit 10 and 20 $\mu$m silicate emission features with varying 
degrees of prominence. Two notable exceptions are the K0-type star [PZ99]J161411.0-230536 and the K2-type star
[PZ99]J160421.7-213028, which shows a steep decline in mid-infrared flux from
$\sim$2.2--16 $\mu$m where a turnover occurs in the SED. Aside from this inflection point, the mid-infrared 
SED of [PZ99]J160421.7-213028 is near featureless. Other than [PZ99]J160421.7-213028, the MIPS 70 $\mu$m fluxes 
of the late-type Upper Scorpius members are in decline. In later spectral types (M3--M5), there is a tentative
detection of the vibration-rotation band of C$_{2}$H$_{2}$ (acetylene) at 13.7 $\mu$m. The feature appears most 
prominently in the M5-type star, J160532.1-193315, but is weakly present in all of the other spectra. The 
14.0 $\mu$m band of HCN (hydrogen cyanide) may be blended with the C$_{2}$H$_{2}$ feature as both have
been identified in low-resolution IRS spectra of young, low-mass stars (Pascucci et al. 2009).

The differences in the SEDs of the Upper Scorpius sample as a function of spectral type are evident in
Figure 5, which shows the median SED for the early (B9--A9), near solar-mass (K5--M2), and late-type (M3--M5)
excess members. The median SEDs from 1.25 to 34.0 $\mu$m were determined  using
extinction-corrected 2MASS photometry and the IRS spectra, which were integrated over identical narrow 
passbands and divided by the width of each band in microns. These photometric measurements were
centered at: 5.7, 7.1, 8.0, 9.2, 9.8, 11.3, 12.3, 13.25, 16.25, 18.0, 21.0, 25.0, 30.0, and 34.0 $\mu$m.
The adopted bandwidths were 0.6 $\mu$m for wavelengths $\le16.25$ $\mu$m, 2.0 $\mu$m for the bands centered
near the silicate emission features (9.8 and 18.0 $\mu$m), and 3.0 $\mu$m for the longer wavelengths.
For consistency with the Furlan et al. (2006) analysis of young stars in Taurus-Auriga, all fluxes were normalized at $H-$band prior
to computing the median SED. Upper and lower quartiles were also computed for the median SEDs,
which define the range of flux levels for 50\% of the stars in each spectral type bin. Superposed in Figure 5
are the NextGen stellar atmospheric models of
Hauschildt et al. (1999), fitted near $H-$band, for $T_{\rm eff}$ of 9600 K (A0), 3800 K (M0),  and 3200 K (M5),
respectively. Emission in excess of the stellar photosphere first becomes apparent near 8 $\mu$m for the 
early-type stars, between 2.2 and 4.5 $\mu$m for the near solar-mass stars, and near 4.5 $\mu$m for the low-mass excess 
members. The 10 $\mu$m solid state emission feature appears strongest in the solar analog (K5--M2) population,
significantly less prominent in the low-mass (M3--M5) stars, and very weak or absent altogether in the early-type 
(B9--A9) SED. The median SEDs for each range of spectral type and their upper and lower quartiles are 
presented in Table 2.

\subsection{Notes on Individual Sources}

[PZ99]J160421.7-213028: The SED of this K2-type star exhibits a steep decline from $\sim$2.2 to 16 $\mu$m
where a sharp turnover occurs. There is, however, significant disparity (factor of 4) between the measured
IRAC 4.5 and 8.0 $\mu$m fluxes and the flux level of the IRS spectrum. The IRS 16 $\mu$m photometric point
(obtained 10 days before the IRAC photometry) agrees well with the much later IRS spectrum and MIPS photometry.
Comparing the 2MASS and DENIS photometry, we find that the $J-$ and $K-$band magnitudes vary by 0.47 and 0.24 
mag, respectively. A fainter field star 16\arcsec\ distant from
[PZ99]J160421.7-213028 exhibits $J-$ and $K-$band magnitude differences of 0.01 and 0.02 mag, respectively.
The implication is that [PZ99]J160421.7-213028 is a variable star, but the origin of its variability is unknown.
The SED of [PZ99]J160421.7-213028 is similar to that of the K2-type
weak-line T Tauri star, UX Tau A in the Taurus-Auriga molecular cloud complex. Both lack significant silicate 
emission features, but whereas the SED of UX Tau A plateaus from 20 to $\sim$100 $\mu$m, that of [PZ99]J160421.7-213028 
continues to sharply rise to at least 70 $\mu$m. UX Tau A is also weakly accreting with $\dot{M}\sim10^{-9}$ M$_{\odot}$ yr$^{-1}$. 
Espaillat et al. (2007) model the SED of UX Tau A as a 2-component disk system: an optically thick inner disk 
of gas and large dust grains and an optically thick outer disk truncated near $\sim$56 AU. The inner disk of
[PZ99]J160421.7-213028 is likely depleted of micron-sized dust grains and gas given the lack of solid state
emission features and accretion indicators (see \S 4.2 and 4.3).

J160643.8-190805: This K6-type WTTS exhibits near photospheric emission to $\sim$8 $\mu$m, where a slight
excess first becomes apparent. The SED suggests that the inner AU of disk structure is devoid of significant
quantities of dust. The 10 and 20 $\mu$m silicate emission features are extremely weak and significantly
broadened. The star was not detected in the MIPS 70 $\mu$m survey implying that excess emission remains weak
into the far infrared. J160643.8-190805 is not accreting based upon the low velocity width of H$\alpha$ emission 
(\S 4.2 and 4.3).

J161420.2-190648: This M0-type star exhibits the largest infrared excess of all disk-bearing members in 
the Upper Scorpius sample. The SED shown in Figure 1c is normalized near $I-$band given the sharply rising near-infrared
($JHK_{S}$) flux levels. Excess emission is evident from $J-$band and extends redward to at least 70 $\mu$m.
The SED of the star is reminiscent of the Class II sources in Taurus-Auriga with prominent 10 and 20 $\mu$m 
silicate emission peaks. The MIPS 70 $\mu$m flux level suggests that the SED declines redward of the $\sim$20 $\mu$m 
solid state feature. The mass accretion rate for this star is the highest of all suspected accretors in the 
Upper Scorpius sample, $\dot{M}=$10$^{-8.9}$ M$_{\odot}$ yr$^{-1}$ (\S 4.2 and 4.3).

J161115.3-175721: The SED of this M1-type star exhibits near-photospheric emission in the near-infrared ($JHK_{S}$).
By 4.5 $\mu$m, however, excess emission is readily apparent and continues to at least 38 $\mu$m, the limit of the 
IRS spectral range. The 10 $\mu$m solid state emission feature is very prominent, but redward of the 
20 $\mu$m silicate emission peak, the flux level rapidly declines, possibly the result of a truncated outer disk.
The star was not detected in the MIPS 70 $\mu$m survey, suggesting that the low level of 
excess emission evident at the limit of the IRS wavelength coverage continues beyond. This star is not 
suspected of accretion activity.

J160959.4-180009: This M4-type WTTS exhibits a near-photospheric SED out to 4.5 $\mu$m where excess
emission first becomes apparent. The 10 and 20 $\mu$m solid state features are among the
strongest found in the sample. J160959.4-180009 was the only late-type (M3--M5) excess member detected
at 70 $\mu$m. Although not suspected of accretion activity (\S 4.2 and 4.3), [O I] $\lambda$6300 is detected
in emission, possibly associated with a stellar wind.

J160532.1-193315: This accreting M5 star has a peculiar mid-infrared spectrum that exhibits broad
emission peaks centered near $\sim$7.7 and 13.7 $\mu$m. Polycyclic aromatic hydrocarbon (PAH) emission,
possibly excited by the hot accretion flux, may be responsible for the broad 7.7 $\mu$m feature,
but the 13.7 $\mu$m band is tentatively identified as the vibration-rotation band of C$_{2}$H$_{2}$ 
(Pascucci et al 2009). Other possible progenitors include amorphous water ice,
PAH emission, or some combination of these sources. Alternatively, the SED of J160532.1-193315
may be interpreted as exhibiting shallow silicate absorption at 10 and 20 $\mu$m, resulting in the
observed ``peaks'' between the absorption features. Broad 10 $\mu$m silicate absorption is commonly
observed in embedded class I sources experiencing infall (e.g. Zasowski et al. 2008) or in edge-on
disk systems where cold dust obscures the line of sight. At 20 $\mu$m, however, the SEDs of young
protostars are dominated by rising continuum emission, which effectively veils absorption from silicates.
For J160532.1-193315 this interpretation is difficult given the substantial rise above the continuum
level from 7.0--7.7 $\mu$m before turning over. The depth of the proposed 10 $\mu$m absorption feature
also coincides with the expected continuum level if one extrapolates the SED through the enhancement
using the slope from 5.2--7.0 $\mu$m.

\section{Accretion Diagnostics}

In the magnetospheric accretion model, gas from the inner disk edge is channeled along lines of
magnetic flux to the stellar surface where its impact upon the photosphere generates hot continuum
excess emission with peak temperatures of $\sim$8,000--12,000 K (Valenti et al. 1993;
Hartman et al. 1998; Muzerolle et al. 1998). The infalling gas is inferred from inverse P Cygni
line profiles as well as broadened \ion{H}{1}, \ion{He}{1}, and \ion{Ca}{2} emission lines with
velocity widths often exceeding several 100 km s$^{-1}$ (Hamann \& Persson 1992; Batalha \& Basri 1993).
The mass accretion rate, $\dot{M}$, is most reliably determined using blue spectrophotometry to measure the hot continuum 
excess emission directly from the Balmer jump (Valenti et al. 1993; Gullbring et al. 1998; Herczeg \& Hillenbrand 2008), 
or by using dereddened $U-$band photometry, which correlates strongly with accretion luminosity (Gullbring et al. 1998).
In the absence of available blue spectrophotometry, reliable accretion diagnostics have been developed
in the red for use in extincted regions. Continuum excess emission or optical veiling near
$\lambda$6500 allows a direct measurement of accretion luminosity, $L_{acc}$, to be made using 
high resolution optical spectra. The extinction-corrected \ion{Ca}{2} $\lambda$8542 line luminosity 
also has a well-established linear relationship with $L_{acc}$ (Muzerolle et al. 1998; 
Herczeg \& Hillenbrand 2008; Dahm 2008) that permits an independent estimate of $L_{acc}$. The derived 
$L_{acc}$ values can then be transformed to $\dot{M}$ from the relation:

$L_{acc} \sim \frac{G  M_{*} \dot{M}}{R_{*}} (1 - \frac{R_{*}}{R_{in}})$ (1)

\noindent where the stellar mass and radius estimates are obtained from the evolutionary models of
Siess et al. (2000). The factor $(1 - R_{*}/R_{in})$ is assigned a value of 0.8, which assumes an
inner disk radius $(R_{in})$ of 5 $R_{*}$ (Gullbring et al. 1998). Given the possibility of larger
inner disk radii for the excess members of Upper Scorpius, this value could be underestimated by
a factor of 1.25.

\subsection{H$\alpha$ Emission}
White \& Basri (2003) use the velocity width of H$\alpha$ at 10\% peak flux to distinguish between
optically veiled and non-veiled stars, i.e. between suspected accretors and non-accretors. Their
velocity width criterion of 270 km s$^{-1}$ is independent of spectral type and of uncertainties
introduced when adopting photospheric template spectra for the veiling analysis. The White \& Basri (2003)
H$\alpha$ velocity width criterion was applied to the 35 excess members of Upper Scorpius observed
with HIRES, identifying only 7 suspected accretors. The M5 star J160827.5-194904 exhibits an H$\alpha$
velocity width of $\sim$255 km s$^{-1}$, however, strict adherence to the White \& Basri (2003) accretion 
criterion is maintained. All other non-accretors have H$\alpha$ velocity widths $<$ 225 km s$^{-1}$
with a median value of 131 km s$^{-1}$. The presence of [O I] $\lambda$6300 and [S II]
$\lambda$6731 emission, indicators of accretion shocks and stellar or disk winds, were also used to
support the accretion analysis. Shown in Figure 6 are the  H$\alpha$ emission profiles for the
7 suspected accretors in the Upper Scorpius disk sample. The strongest accretor (i.e. possessing
the highest $\dot{M}$) is the M0-type star, J161420.2-190648, which exhibits significant infrared 
excess emission from $J-$band redward.
Among the low-mass excess members of Upper Scorpius, the earliest disk-bearing star is the K0-type
[PZ99]J161411.0-230536, which exhibits infrared excess emission at 4.5, 8.0, and 16.0 $\mu$m.
An HIRES spectrum of this star obtained in 2008 June suggests
that accretion may be occurring. H$\alpha$ exhibits an inverse P Cygni line profile indicative 
of mass infall, but only weak emission is present. Integrated over the full line profile,
the equivalent width of H$\alpha$ is W(H$\alpha$)$=+$0.38 \AA. No forbidden emission lines 
generally associated with protostellar jets or accretion shocks are evident. The earliest definitive 
accretor in the Upper Scorpius disk sample is the K5-type star [PZ99]J160357.6-203105, in which 
non-photospheric emission first becomes detectable between $K-$band and 4.5 $\mu$m.

\subsection{Continuum Excess Emission}
Veiling or continuum excess emission, defined such that $r = F_{exc}/F_{phot}$, was determined for
the 7 suspected accretors by fitting the depths of several \ion{Ca}{1} and \ion{Fe}{1} photospheric
absorption lines from 6000--6500\AA\ with those of a rotationally broadened and artificially veiled
standard star of similar spectral type (White \& Hillenbrand 2004). Veiling errors are dominated
by the uncertainty in spectral type and are estimated to be $\pm$0.05--0.1 for lightly veiled stars 
and up to $\pm$0.3 for the most heavily veiled object. To estimate the continuum excess luminosity, 
the stellar photospheric flux was determined using the extinction-corrected $R-$band mag.
To transform the excess fluxes to total
accretion luminosity, a multiplicative bolometric correction was applied to account for the
accretion-generated flux outside of the $R$ filter bandpass. We adopt a bolometric correction of
11.1, following the reasoning of White \& Hillenbrand (2004), which is the logarithmic average
of the 2 values proposed by Gullbring et al. (1998), 3.5, and Hartigan \& Kenyon (2003), 35.
The mass accretion rate then follows using the stellar mass and radius estimates of the
Siess et al. (2000) pre-main sequence models. The veiling estimates for the 7 accretors
in Upper Scorpius range from near zero up to $\sim$ 1.5 for the M3-type star J155829.8-231007.
The resulting $\dot{M}$ values range from $10^{-8.9}$ to $10^{-11.0}$ M$_{\odot}$ yr$^{-1}$ and 
are listed in Table 3. Uncertainties for these estimates include errors in continuum excess
emission (factor of $\sim$2), the adopted bolometric correction (factor of $\sim$3), and the 
estimate for the inner disk radius (factor of $\sim$1.25). We therefore assign uncertainties
of 0.5--1.0 dex for the derived log $\dot{M}$ values presented in Table 3.

\subsection{\ion{Ca}{2} $\lambda$8542 Emission}
In their examination of accreting CTTS in the Taurus-Auriga star forming region, Muzerolle et al. (1998)
find that the \ion{Ca}{2} $\lambda$8542 emission line luminosity is a strong correlate of $\dot{M}$.
The linear relation persists through at least 3 orders of magnitude from $\dot{M}$$\sim$10$^{-6.0}$ to
10$^{-9.0}$ M$_{\odot}$ yr$^{-1}$. Dahm (2008) used published and observed \ion{Ca}{2} $\lambda$8542
emission line luminosities for a dozen accreting CTTS in Taurus and their $L_{acc}$ values obtained
from the blue continuum excess measurements of Valenti et al. (1993) and Gullbring et al. (1998) to
derive a linear regression fit given by:

log$(L_{acc}/L_{\odot})=(0.94\pm0.11)$log$(L_{\lambda8542}/L_{\odot})+(2.64\pm0.38)$ (2)

\noindent Some scatter in the relationship exists, likely resulting from the non-simultaneous nature of the \ion{Ca}{2} and
blue continuum spectrophotometric observations. Herczeg \& Hillenbrand (2008) find a similar linear
fit for accreting low mass stars and substellar objects in their {\it simultaneous} blue continuum
excess measurements and \ion{Ca}{2} $\lambda$8542 emission line luminosity determinations:

log$(L_{acc}/L_{\odot})=(1.02\pm0.11)$log$(L_{\lambda8542}/L_{\odot})+(2.50\pm0.5)$ (3)

\noindent To measure $L_{\lambda8542}$, the \ion{Ca}{2} $\lambda$8542 absorption line profile from 
the normalized spectrum of a standard star of similar spectral type was subtracted from the profile
of each suspected accretor. The emission line luminosity was then estimated using the measured
equivalent width, the extinction-corrected $I-$band mag, the spectral type dependent flux ratio
between $\lambda$8542 and the effective wavelength of the $I-$band filter (obtained from the 
spectrophotometry of O'Connell 1973), and by assuming a distance of 145 pc for the Upper Scorpius
OB associtation. $L_{acc}$ estimates were then determined using the above
relationships from Dahm (2008) and Herczeg \& Hillenbrand (2008). In general, the linear relation
of Dahm (2008) tends to overestimate $L_{acc}$ and therefore $\dot{M}$ relative to the continuum
excess $\dot{M}$ values (by up to an order of magnitude). We adopt the relationship of
Herczeg \& Hillenbrand (2008) given their simultaneous and independent determinations of $L_{acc}$
and $L_{\lambda8542}$ and the lower stellar masses sampled by their investigation. The $\dot{M}$
values agree well (factor of $\sim$2--3) with those from the continuum excess analysis and are provided 
in Table 3.

\section{Comparison of the Upper Scorpius and Taurus-Auriga Disk Samples}

The inner disk dissipates in $\sim$50\% of low mass stars by an age of $\sim$3\,Myr (Haisch et al. 2001). 
Consistent with this trend, $\sim$50\% of stars in Taurus-Auriga (age $\sim$ 1-3\,Myr) have inner 
circumstellar disks (Kenyon \& Hartman 1995), while only 20\% of stars in Upper Scorpius (age $\sim$5\,Myr) 
have inner disks (Carpenter et al. 2006). We now examine the SEDs of these two populations to establish if the presumed evolutionary 
changes in the percentage of stars with an infrared excess is also accompanied by changes in the disk 
structure, as traced by variations in the SED shape and accretion diagnostics. Since Taurus-Auriga is a low-mass star forming 
region, we focus the analysis on the K and M-type members of each region. We compare in turn the infrared 
excesses between 2.2 and 8 $\mu$m, various mid-infrared spectral indices derived from the IRS spectra, and 
mass accretion rates.

\subsection{The Taurus-Auriga Class II and Class III Samples}
Furlan et al. (2006) present {\it Spitzer} mid-infrared spectra for 85 Class II and 26 Class III objects 
in the Taurus-Auriga star forming region. Early classification of Taurus sources (e.g. Kenyon \& Hartmann 1995)
was based upon the slope ($\alpha$) of the SED from 2.2--25 $\mu$m, defined such that 
$\alpha = \delta$log($\lambda F_{\lambda}$)/$\delta$log($\lambda$).
Sources having SED slopes of $-2 < \alpha < 0$ were defined as Class II objects which exhibit infrared excesses 
characteristic of circumstellar disk emission. Sources with $\alpha < -2$ were defined as Class III objects which 
exhibit no detectable infrared excess and have SEDs characteristic of pure stellar photospheres. This classification 
scheme is critically dependent upon IRAS photometry and does not account for the detailed shape of the mid-infrared SED.
Consequently, Furlan et al. (2006) re-classified the Taurus-Auriga sources based upon the detection of excess emission 
in the IRS spectral range (5.2--38 $\mu$m). Given the significantly greater sensitivity of {\it Spitzer},
several objects previously identified as Class III sources using the {\it IRAS} 25 $\mu$m measurements, 
were reclassified as Class II by Furlan et al. (2006). The stars were also divided into classical and weak-line 
T Tauri stars (CTTS/WTTS) using the modified, spectral-type dependent accretion criterion of White \& Basri (2003). 
Among the sample of Class II sources in Taurus, Furlan et al. (2006) find that most are CTTS and suggest that 
inner disk gas and dust are responsible for a substantial fraction of the infrared excess. The Upper Scorpius 
disk-bearing sample is similar in many respects to the redefined Class II population of Taurus-Auriga. Both 
samples were classified on the basis of emission excesses within the IRS spectral range and cover similar spectral 
types ($\sim$B8--M5). All of the Upper Scorpius excess members would be classified as Class II sources by the
Furlan et al. (2006) definition. Although neither sample is complete, both can be regarded as representative 
of the larger disk-bearing populations of each region.

\subsection{Comparison of Near and Mid-Infrared Photometry}
In Figure 7, we present the 2MASS $H-K_{S}$, $J-H$ color-color diagram for the Upper Scorpius and Taurus-Auriga
disk-bearing samples. The 35 Upper Scorpius members appear clustered around the locus of main sequence stars, with 
the early-type members lying near the base of the trunk and the late-type stars just above the M-dwarf branch. There 
is an obvious lack of significant reddening toward Upper Scorpius as the largest extinction is $A_{V}$ $\sim$3 mag.
In contrast to the Upper Scorpius disk-bearing population, the 71 Taurus-Auriga Class II sources from Furlan et al. (2006)
with 2MASS photometry available suffer significant reddening (either interstellar in origin or local extinction
effects), and over half exhibit colors that place them outside of the reddening boundaries 
for normal stars. The Class III sources of Taurus-Auriga lie clustered around the dwarf locus or along its reddening 
vector, demonstrating a lack of infrared excess emission. 

To quantify the near-infrared excesses for these two populations, we show in Figure 8 the $E(H-K_{S})$ histogram for 
Upper Scorpius disk-bearing stars (shaded region) and the Taurus-Auriga Class II sources (open region). Given
that the early-type members of Upper Scorpius have no direct counterparts in Taurus-Auriga, they are not included
in the histogram sample. Extinction corrections are made using the $A_{V}$ values from Furlan et al. (2006) for 
members of Taurus-Auriga and those of Preibisch \& Zinnecker (1999) or Preibisch et al. (2002) for members of 
Upper Scorpius. To determine color excesses, intrinsic 2MASS $H-K_{S}$ colors were obtained from a tabulation 
of B2--K5 main sequence colors adopted by the FEPS survey (Carpenter et al. 2008). Intrinsic 2MASS $H-K_{S}$ 
colors for later spectral types (K5--M5) were derived using $\sim$1100 classified stars from the Palomar/MSU Nearby
Star Spectroscopic Survey of Reid et al. (1995) and Hawley et al. (1996). Given their proximity to the Sun, 
the stars suffer little extinction. Sample sizes for each half-spectral class bin range from $\sim$20 to 160,
thereby ensuring statistical robustness. The median $H-K_{S}$ color for each bin is adopted as the intrinsic 
color.

The distributions of $H-K_{S}$ color-excesses are clearly different, with median $E(H-K_{S})$ values of 0.03 and 
0.35 for the Upper Scorpius and Taurus-Auriga samples, respectively. The small $H-K_{S}$ excess observed in Upper 
Scorpius could result if the intrinsic $H-K_{S}$ colors are dependent on surface gravity; the Upper Scorpius
members are systematically younger than the K- and M-type stars used to derive the intrinsic $H-K_{S}$ colors,
and as a result have lower surface gravities. Thus only two of the Upper Scorpius sources (J161420.2-190648
and [PZ99]J160421.7-213028) have clear $H-K_{S}$ excesses (see Fig. 8).
We find that hot dust grains within the inner disk region 
($\ll$1 AU) are more prevalent among the presumably younger Taurus-Auriga Class II sources.
Such near-infrared excess originates from reprocessed stellar radiation emitted by the thermally flared inner disk 
rim at or near the evaporation temperature of dust, $\sim$1500 K. One consequence of the flared inner disk rim
is that the interior region is shadowed, causing a decrease in scale height and a reduction of infrared emission
(Dullemond et al. 2001). Beyond the shadowed region, the scale height of the disk increases  due to the reduced 
gravitational potential. Consequently, the surface layer of the disk is again exposed to stellar radiation. 
The models of Dullemond et al. (2001) predict that emission from this surface layer and the disk interior peak 
at wavelengths in the near and mid-infrared, respectively. 

Hartmann et al. (2005), Luhman et al. (2006), and Padgett et al. (2008) present {\it Spitzer} IRAC
3.6, 4.5, 5.8, and 8.0 $\mu$m photometry for 56 Class II and III sources in the Furlan et al. (2006) 
Taurus-Auriga sample. Complementing these data are the IRAC 4.5 and 8.0 $\mu$m fluxes from 
Carpenter et al. (2006) for all infrared excess members of Upper Scorpius. The IRAC fluxes were converted to magnitudes 
using the zero magnitude flux densities of 179.7 Jy for channel 2 and 64.13 Jy for channel 4 (Reach et al. 2005). 
Shown in Figures 9 and 10 are the histograms of the $E(K_{S}-[4.5])$ and $E(K_{S}-[8.0])$ color excess 
for the 24 low-mass disk-bearing members of Upper Scorpius and the 45 Taurus-Auriga Class II 
sources of Furlan et al. (2006) with IRAC photometry available. To obtain intrinsic $K_{S}-[4.5]$ and $K_{S}-[8.0]$
colors for the Taurus-Auriga and Upper Scorpius samples, stellar photospheric 4.5 and 8.0 $\mu$m fluxes were 
estimated using the {\it Spitzer} Stellar Performance Estimation Tool (STAR-PET), which adopts the Kurucz-Lejeune 
stellar atmospheric models. The observed $K_{S}$ magnitudes (from 2MASS) are then corrected for extinction using 
the $A_{V}$ values provided by Furlan et al. (2006) for Taurus-Auriga and by Preibisch \& Zinnecker (1999)
or Preibisch et al. (2002) for members of Upper Scorpius. The $E(K_{S}-[4.5])$ distributions are markedly
different with median color excesses of 1.37 and 0.77 mag for the Taurus-Auriga and Upper Scorpius samples, 
respectively. The $E(K_{S}-[8.0])$ distributions for these samples have median values of 2.23 and 1.43 mag 
for Taurus-Auriga and Upper Scorpius, respectively. The differences in color excesses unambiguously 
demonstrate the reduced levels of 2.2--8.0 $\mu$m excess emission in the Upper Scorpius sample relative 
to the Taurus-Auriga Class II sources.

\subsection{Comparison of Mid-Infrared Spectral Indices}
The 2MASS and IRAC photometry reveal clear differences in the amount of warm dust in the inner disk
between the Upper Scorpius and Taurus-Auriga samples. We now use the IRS spectra to investigate the
differences in the excess characteristics between 8.0 and 38 $\mu$m, which probes dust emission at 
larger orbital radii. To characterize the SEDs for the disk-bearing members of Upper Scorpius 
we determine spectral indices that sample mid-infrared continuum emission.
Following the example of Furlan et al. (2006), we integrate the observed flux in three passbands
defined from 5.4--6.0 $\mu$m, 12.5--14.0 $\mu$m, and 23.5--26.5 $\mu$m and then divide the resulting fluxes
by the width of the passband in microns. The spectral indices are then determined from:

$n = log(\frac{\lambda_{2} F_{\lambda_{2}}}{\lambda_{1} F_{\lambda_{1}}}) / log(\frac{\lambda_{2}}{\lambda_{1}})$ (4)

\noindent In this manner, 3 mid-infrared spectral indices are determined, from 5.7 to 13.25 $\mu$m ($n_{6-13}$), from 
13.25 to 25 $\mu$m ($n_{13-25}$), and from 5.7 to 25 $\mu$m ($n_{6-25}$). An additional index measuring 
the strength of the 10 $\mu$m silicate emission feature is also determined by interpolating the continuum 
from 5.2--7.0 $\mu$m and from 13.0--16.0 $\mu$m with a fifth or sixth order polynomial. The resulting fitted 
continuum was then subtracted from the silicate emission peak defined between 8.0 and 12.5 $\mu$m. The
integrated flux of the continuum-subtracted silicate feature is then normalized to the fitted continuum level 
between 8.0 and 12.5 $\mu$m. Table 4 tabulates the spectral indices and the 10 $\mu$m silicate emission
strengths for the Upper Scorpius members.

In Figure 11 we present histograms of the $n_{6-13}$ index for the low-mass (K0--M5) members of
Upper Scorpius and the Class II population in Taurus-Auriga. The median value of the $n_{6-13}$ 
index is $-0.87$ for the low-mass stars in Upper Scorpius, and $-0.79$ for the Class II population 
of Taurus-Auriga. The Taurus-Auriga sample tends to have more outliers with positive indices, while
the Upper Scorpius sample has more sources with negative indices. Nonetheless, the Kolomogorov-Smirnov
statistic ($d$) for these samples is $d=0.19$ with a significance level of 0.95, suggesting  that the 
two samples are consistent with having been drawn from the same parent population. Similarly, in
Figure 12 we present histograms of the $n_{13-25}$ indices. The median values of the $n_{13-25}$ 
distributions are $-0.35$ and $-0.14$ for Upper Scorpius and Taurus-Auruga, respectively. However,
the Kolomogorov-Smirnov statistic for these two samples is $d=0.23$ with a significance level of 0.78 
and suggests that the distributions of $n_{13-25}$ indices are indistinguishable.

Shown in Figure 13 is the $n_{13-25}$ mid-infrared spectral index plotted against the $n_{6-13}$
index for the 26 excess members of Upper Scorpius observed with IRS and the 85 Class II and 26
Class III sources in Taurus-Auriga from Furlan et al. (2006). The early-type excess members of 
Upper Scorpius appear well displaced from the low-mass stars and lie between the Class II and III 
populations of Taurus-Auriga. The oddly placed low-mass Upper Scorpius member near 
$n_{6-13}$$\sim$$-$0.75, $n_{13-25}$$\sim$$-$2.8 is the M5-type star J160532.1-193315, which exhibits
possible PAH and acetylene emission features near 7.7 and 13.7 $\mu$m (\S 3.2). Overall, the two dimensional
distributions of the $n_{6-13}$ and $n_{13-25}$ indices are similar for the low-mass disk-bearing 
stars in the Upper Scorpius and Taurus-Auriga samples.

Furlan et al. (2006) define a morphological sequence (labeled A--E) of mid-infrared spectra using silicate 
emission strengths and the slope of the mid-infrared SED as classification criteria. This classification
scheme may be representative of a simple disk evolutionary sequence such that as micron-sized dust grains 
are depleted, 10 $\mu$m silicate emission broadens and weakens resulting in a more negative SED slope.
Shown in Figure 14 is the continuum-subtracted, integrated flux of the 10 $\mu$m silicate emission feature,
normalized to the continuum and plotted as a function of the $n_{6-25}$ spectral index for the 
Upper Scorpius excess members observed with IRS. Also shown in the figure are the 85 Class II sources in 
Taurus-Auriga and the approximate boundaries of the various classes (A--E) in the morphological sequence 
defined by Furlan et al. (2006). Taurus-Auriga sources lying outside of these boundaries were identified 
as outliers of the sequence. In general, for both populations, stars with $n_{6-25}$ $< -1.0$ exhibit weak
10 $\mu$m silicate emission. Considering just the low-mass Upper Scorpius excess members, we find
the peak-to-peak range of the 10 $\mu$m silicate emission index to be somewhat higher for the 
Taurus-Auriga sample, but given its larger size there is a greater probability for the presence of outliers.
The ranges of $n_{6-25}$ indices are comparable for both samples, however, the distribution is
shifted toward more negative values for the Upper Scorpius sample. None of the Taurus-Auriga Class II sources 
are found with $n_{6-25}$ indices $<$ $-1.2$, whereas 4 (24\%) low-mass Upper Scorpius members 
occupy this region of the diagram.

\subsection{The Median Spectral Energy Distribution for Near Solar Mass Stars}
In previous sections, we compared the Upper Scorpius and Taurus-Auriga low-mass disk-bearing samples 
using selected photometric bands and indices. We now compare the disk properties based on the global shape of the IRS spectra.
The median SED from 1.25 to 34.0 $\mu$m of the 8 Upper Scorpius near solar-mass (K5--M2) excess stars 
is shown in Figure 15 with its upper and lower quartiles superposed. Also shown is the median SED of 
the 55 Taurus-Auriga Class II sources within the same range of spectral types from Furlan et al. (2006).
The 10 $\mu$m silicate emission feature is of similar strength in both median SEDs, however, the 20 $\mu$m
peak is significantly more pronounced in the Upper Scorpius sample. This may result from enhanced optically 
thin 20 $\mu$m emission from the disk surface layer. The slope of the median SED from 1.6 to $\sim$8.0 $\mu$m 
is significantly steeper for the K5--M2 stars in Upper Scorpius relative to the Taurus-Auriga Class II sample.
This provides additional supporting evidence for the reduced levels of inner disk emission inferred from
the near ($K_{S}$, 4.5 $\mu$m) and mid-infrared (8.0 $\mu$m) color excess distributions. 

\subsection{Synthesis}
We now combine the results in \S{5.2-5.4} to summarize the differences and similarities between disks 
around low mass stars in the Upper Scorpius and Taurus-Auriga samples. Not only do a greater fraction 
of stars in Taurus-Auriga possess circumstellar disks than low-mass stars in Upper Scorpius, the disk 
properties are clearly different as well. The most obvious difference is that disks in Upper Scorpius
lack the hot dust grains prevalent around Class II sources in Taurus-Auriga, as evidenced by the reduced 
levels of excess emission at 2.2 $\mu$m (see Figs. 7 and 8), 4.5 $\mu$m (Figs. 9), and 8 $\mu$m (Fig. 10), 
and further exemplified by the sharply declining median SED (Fig. 15). Qualitatively, these properties are
similar to the ``anemic'' disks found in IC\,348 (Lada et al. 2006), although unlike IC\,348, the reduced
level of excesses at $<$ 8 $\mu$m is found in nearly all late-type stars in Upper Scorpius. The ubiquity
of depleted inner disks in Upper Scorpius suggests this is a common evolutionary pathway that persists
for an appreciable fraction of the disk lifetime. Differences in the inner disk properties
also extend to the gaseous component; the mean $\dot{M}$ (10$^{-9.0}$ M$_{\odot}$ yr$^{-1}$) for the 6 
Upper Scorpius accretors with masses between 0.27 and 1.07 M$_{\odot}$ is an order of magnitude lower
than that of 17 accreting CTTS in Taurus-Auriga (10$^{-7.97}$ M$_{\odot}$ yr$^{-1}$) within a similar mass range
(Gullbring et al. 1998). The lower mean $\dot{M}$ value and the clearly different infrared excess properties
of the Upper Scorpius sample relative to the Class II population of Taurus-Auriga imply significant
differences in inner disk structure and composition.

The disk properties can be constrained using the grid of radiative transfer models from Robitaille et al. (2006, 2007).
These models assume that an optically thick accretion disk surrounds a pre-main-sequence star, and the resultant SED 
is computed for various combinations of parameters, including inclination, stellar mass and radius, mass accretion 
rate, inner and outer disk radius, disk mass, distance, and extinction. The grid of model spectra were fitted to 
the SEDs of a representative sample of near solar-mass (K5--M2) and low-mass (M3--M5) Upper Scorpius members to
determine the range of inner disk radii that are consistent with the observations. To better constrain inner
disk emission, only the 2MASS photometry, IRAC 4.5 and 8.0 $\mu$m fluxes from Carpenter et al. (2006), three
narrow photometric bands of the IRS spectra centered at 5.7, 7.1, and 9.2 $\mu$m, and the MIPS 24 $\mu$m fluxes
from Carpenter et al. (2009, in preparation) were used in the model fitting. Distance and extinction were
allowed to vary from 100 to 200 pc and from 0.0 and 3.0 mag, respectively. The inner disk radii for the models 
most consistent with the SEDs of the near solar-mass and low-mass stars ranged from 0.2--0.8 AU and from 0.4--1.2 AU,
respectively. The model comparisons support the conclusion that the inner disk regions of the Upper Scorpius
disk-bearing population are evacuated of dust and gas.

The shape of the IRS spectra at longer wavelengths probes the disk structure at larger orbital radii. 
While we do not find any significant differences in the continuum slope between the disks around stars 
in the Upper Scorpius and Taurus-Auriga samples (Figs. 11 and 12), the amount of infrared emission is 
lower by a factor of $\sim$2 in Upper Scorpius for stars with spectral types between K5--M2 (Fig. 15).
The reduced levels of excess emission are apparent from 2.2--34.0 $\mu$m, with a peak factor of $\sim$3
difference occurring near 8.0 $\mu$m. Assuming a disk temperature distribution given by D'Alessio et al. (1999)
for accretion disks around low-mass stars:

$T_{r} = T_{eff} (\frac{\lambda}{7})^{2/7} r^{-3/7}$ (5)

\noindent where $\lambda$ is a parameter dependent upon stellar mass, radius, and $T_{\rm eff}$, we find
that the IRS spectral range samples the disk from $\ll$1 AU to beyond 20 AU for a typical 0.5 M$_{\odot}$ 
pre-main sequence star. If the changes in the excess properties of the Upper Scorpius disk sample relative
to the Class II population of Taurus-Auriga are interpreted as evidence for disk evolution,
then the disk between $\ll$1 and 20 AU is evolving on 2--4 Myr timescales.

Examination of the 10 and 20 $\mu$m solid state emission features may also provide evidence for inner
disk evolution. As micron-size dust grains in the optically thin disk surface layer coalesce, silicate emission
is expected to broaden and flatten while mid-plane settling decreases the mid-infrared continuum excess emission
(Furlan et al. 2006; D'Alessio et al. 2006). As shown in Figure 5, the strength of silicate emission among
Upper Scorpius excess members is strongly spectral type or mass dependent, with the most prominent features
being associated with the near solar-mass stars. Sicilia-Aguilar et al. (2007) also find a spectral type 
dependence of silicate emission strength, such that weak silicate features are 3--5 times more common in 
M-type stars than in K-type stars in Trumpler 37 and NGC\,7160. They suggest that solid state emission probes 
different regions of the disk for K and M-type stars given that it is produced in closer proximity to the
stellar photosphere and within a smaller disk region for the later spectral types. The Upper Scorpius IRS 
observations support this conclusion, providing additional evidence of the stellar mass dependent nature 
of solid state emission strengths. If differential disk evolution were occurring (e.g. inside-out dissipation),
silicate emission would appear weaker among late M-type stars whose disks may have been cleared beyond the
region probed by such emission. These effects, however, would be subtle and cannot be adequately assessed by
the IRS observations presented here. Relative to the sample of Taurus-Auriga Class II sources, we find
that the 10 $\mu$m silicate emission features among the low-mass members of Upper Scorpius are of comparable
strength. 

\section{Circumstellar Disks Associated with Early-Type Upper Scorpius Members}

The early-type excess members of Upper Scorpius exhibit near-photospheric fluxes out to $\sim$8 $\mu$m 
and weak, featureless continuum excesses redward that extend to the wavelength limit of IRS, suggesting
that the dust grains responsible for the observed excesses are at least several microns in diameter.
This is supported circumstantially by the estimated blowout radius ($a_{rad}$) for dust grains around
these early-type stars due to radiation pressure (Artymowicz 1988):

$a_{rad} = \frac{3 L_{*} Q_{PR}}{16 \pi G M_{*} c \rho}$ (6)

\noindent where $\rho$ is the grain mass density, assumed to be 2.5 g cm$^{-3}$, and $Q_{PR}$ is the radiation
pressure coupling coefficient, assumed to be unity (i.e. the effective cross-section of the grain is approximately
equal to the geometric cross-section, Chen et al. 2006). The blowout radii are found to range from 40.5 $\mu$m
for the B2 V star HIP 77859, to $\sim$1.0 $\mu$m for the A9 V excess member HIP 80088. Grains with smaller radii are
gravitationally unbound and move hyperbolically out of the system over migration timescales that are 3 orders
of magnitude less than the age of the Upper Scorpius OB association assuming the disk is optically thin and gas-poor. 
For these early-type stars, stellar winds may 
also play a crucial role in disk dispersal. Using the relationship for the stellar wind dust blowout radius from
Plavchan et al. (2005) and the wind mass loss rate upper limit ($2\times 10^{-10}$ M$_{\odot}$ yr$^{-1}$)
for main sequence A-type stars from Lanz \& Catala (1992), we estimate the wind dust blowout radii to be
an order of magnitude less than those from radiation pressure alone. While the mass-loss rates for young
stars may be higher, we adopt the latter values to constrain the minimum dust grain sizes around these
early-type mid-infrared excess stars.

To examine the dust properties of the 8 early-type Upper Scorpius members observed with IRS, 
we show the photosphere-subtracted 5--35 $\mu$m spectra in Figure 16. In several of the 
photosphere-subtracted spectra, a slight rise between 10 and 15 $\mu$m is apparent that may
result from a low-level discontinuity between the SL and LL modules. The NextGen atmospheric
models of Hauschildt et al. (1999) were used to estimate the photospheric flux. The error bars represent
the 1-$\sigma$ uncertainties determined when combining the individual IRS spectra. Superposed 
in Figure 16 are the best fits of single-temperature spherical blackbodies (plotted in red) 
to the observed excess emission. Presented in Table 5 are these blackbody temperatures, the dust 
grain blowout radii ($a_{rad}$), the minimum orbital radii of the dust grains ($r_{min}$),
the fractional infrared luminosities ($L_{IR}/L_{*}$) for the stars, and an estimate of the minimum
dust masses ($M_{dust}$) responsible for the observed excess emission for the early-type Upper Scorpius 
debris disks. The minimum orbital radii are estimated using:

$r_{min} = \frac{1}{2}(\frac{T_{\rm eff}}{T_{gr}})^2 R_{*}$ (7)

\noindent where the grain temperature, $T_{gr}$, is assumed to be the blackbody temperature derived 
from the photosphere-subtracted SED. The stellar radius ($R_{*}$) and $T_{eff}$ values are taken from 
the models of Siess et al. (2000). This simple relation for the minimum orbital radii assumes that
the dust grains act as blackbodies in radiative equilibrium with the host star. The derived minimum
radii range from $\sim$11 AU to more than 50 AU. The infrared luminosities ($L_{IR}$) of the early-type
Upper Scorpius members are derived from:

$L_{IR} = 4 \Omega T_{gr}^4 d^2$ (8)

\noindent where $\Omega$ is the solid angle subtended by the grains (derived from the single-temperature
blackbody radius used to model the excess emission), and $d$ is the distance to the star assumed to
be 145 pc. The resulting fractional infrared luminosities ($L_{IR}$/$L_{*}$) are tabulated in Table 5 
and vary from 10$^{-4}$ to 10$^{-6}$, a typical range for debris disk systems (Chen et al. 2006).
To estimate the minimum mass of the dust responsible for the observed excess emission, we assume the 
dust grains have sizes equal to the blowout radii derived above. Following the assumptions outlined
by Chen et al. (2006) and Jura et al. (1998) for the grain shapes, absorption cross-sections, and the 
distribution of dust around the star, the minimum dust masses are given by:

$M_{dust} \ge \frac{16}{3} \pi \frac{L_{IR}}{L_{*}} \rho r_{min}^2 a_{rad}$ (9)

\noindent The derived dust masses vary considerably, from 1.0$\times$10$^{-4}$ M$_{\earth}$ for the B9V-type HIP 80024
to 1.1$\times$10$^{-6}$ M$_{\earth}$ for the A0-type star HIP 79156. These values, however, are comparable
to dust masses estimated around more evolved debris disk systems (Chen et al. 2006).

The lack of short wavelength ($<$ 8 $\mu$m) excesses, the low fractional infrared luminosities, the
absence of prominent silicate features, and a lack of accretion activity (see \S4) for the A and B-type
stars in Upper Scorpius suggests that the dust emission originates from a debris disk for most sources. 
Note, however, that two of the early-type members (Table 1) are possibly Be-stars and the infrared 
excess likely originates from free-free or free-bound emission from a thin equatorial disk of gas.
The detected 8--10 $\mu$m excesses for {\it all} early-type members of Upper Scorpius is somewhat unusual given that
Chen et al. (2006) find such short wavelength disk emission only among some of their IRAS-selected main 
sequence debris disk sample. Sicilia-Aguilar et al. (2007), however, find similar 8 $\mu$m excesses
associated with several intermediate mass (A0--F5.5) stars in the $\sim$4\,Myr old cluster Trumpler 37 
and the $\sim$10\,Myr old cluster NGC\,7160. If the Upper Scorpius early-type stars are debris disk
systems, they are among the youngest known. A more detailed analysis of the early-type excess members is 
forthcoming (Carpenter et al. in preparation).

\section{Conclusions}

We have examined mid-infrared spectra and high resolution optical spectra of a substantial fraction (26/35) 
of the infrared excess members of the Upper Scorpius OB association identified by Carpenter et al. (2006).
The IRS sample spans spectral types from B8 to M5. This coeval sample of stars, having a well-established
age of $\sim$5\,Myr, is directly compared with the Class II population of the 1--3 Myr old Taurus-Auriga
star forming region. The principal results from the investigation are summarized below.

1. Few (7/35) members of the Upper Scorpius excess population, all late-type (K+M), are found to be accreting using the
H$\alpha$ velocity width accretion criterion of White \& Basri (2003). The mean $\dot{M}$ value
for the 6 Upper Scorpius accretors with masses ranging between 0.27 and 1.07 M$_{\odot}$ (10$^{-9.0}$ M$_{\odot}$ yr$^{-1}$)
is an order of magnitude lower than a similar sample of accreting CTTS in the Taurus-Auriga star
forming region. 

2. Among early-type (B+A) excess stars, the IRS spectra show a featureless continuum with an infrared
excess present for $\lambda \ge 8\mu$m. The lack of 10 and 20 $\mu$m silicate solid state features
indicates that hot sub-micron grains are not present. None of the early-type stars examined with HIRES
show signs of gas accretion. Outside of two likely Be-stars, these properties are qualitatively consistent 
with the properties of debris disk systems.

3. In contrast with the early-type excess stars, all but one late-type (K+M) disk-bearing member of Upper
Scorpius exhibit the 10 and 20 $\mu$m silicate emission features produced by sub-micron scale dust grains 
in the optically thin disk surface layer. A broad range of solid state emission strengths is present,
comparable in many respects to those observed among the Class II population of Taurus-Auriga.

4. We find significant differences in the near-infrared and mid-infrared excess properties between Upper Scorpius
low-mass stars and the Class II population in Taurus-Auriga. Approximately half of the Taurus-Auriga 
Class II sources exhibit $K-$band excesses, but only 2 late-type Upper Scorpius members exhibit
similar excess emission. Moreover, the Upper Scorpius sample shows reduced levels of excess emission 
as measured by the $E(H-K_{S})$, $E(K_{S}-[4.5])$, and $E(K_{S}-[8.0])$ color excess distributions.
Comparing the SEDs of a representative sample of near solar-mass (K5--M2 type) and low-mass (M3--M5 type) 
Upper Scorpius members with radiative transfer models of young stellar objects, we find that the
observed SEDs are consistent with models having inner disk radii ranging from $\sim$0.2 to 1.2 AU.
Between 8 and 35 $\mu$m, the excess flux around K5--M2 stars in Upper Scorpius is a factor of $\sim$2
lower than corresponding stars in Taurus-Auriga. While low-mass stars in Upper Scorpius show reduced
levels of excess emission relative to Taurus-Auriga, the SED slope between 6 and 25 $\mu$m is similar
as measured by the $n_{6-13}$ and $n_{13-25}$ spectral indices. These results suggest that disk evolution
is discernable over the 2--4 Myr in age separating these disk-bearing populations. The ubiquity of 
depleted inner disks in the Upper Scorpius excess sample suggests that such disks are a common evolutionary
pathway that persists for some time.

\acknowledgments
This work is based on observations made with the {\it Spitzer} Space Telescope, which is operated by the Jet Propulsion Laboratory
(JPL), California Institute of Technology, under NASA contract 1407. We have made use of the Digitized Sky Surveys, which were 
produced at the Space Telescope Science Institute under U.S. Government grant NAG W-2166, the SIMBAD database operated at CDS,
Strasbourg, France, and the Two Micron All Sky Survey (2MASS), a joint project of the University of Massachusetts and the 
Infrared Processing and Analysis Center (IPAC)/California Institute of Technology, funded by NASA and the National Science Foundation. 
SED was partially supported by an NSF Astronomy and Astrophysics Postdoctoral Fellowship under award AST-0502381.

\clearpage
\renewcommand{\thefigure}{\arabic{figure}\alph{subfigure}}
\setcounter{subfigure}{1}
\begin{figure}
\epsscale{0.5}
\hspace{2cm}  \vspace{2cm}  \includegraphics[width=11cm,angle=0]{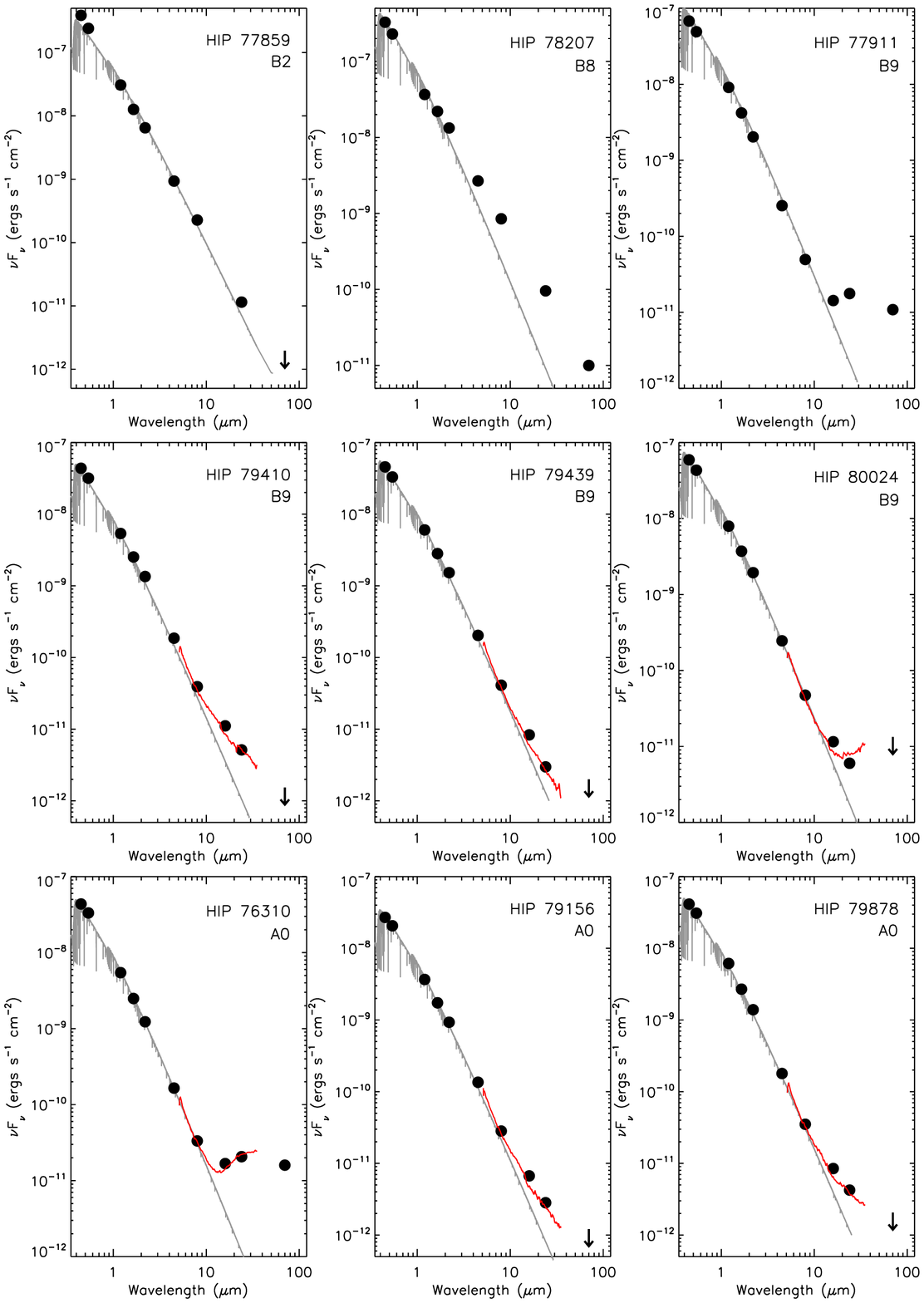} \vspace{-2cm} 
\caption[f1a.ps]{a) Spectral energy distributions for the 35 disk-bearing members of the Upper Scorpius OB
association, ordered by spectral type (early to late) and increasing right ascension. The IRS mid-infrared
spectra (if available) are boxcar smoothed and plotted in red. Also shown are the extinction corrected
optical ($BVRI$) and near-infrared ($JHK_{S}$) photometric data from the literature and 2MASS, {\it Spitzer} IRAC 4.5
and 8.0 $\mu$m and the IRS 16 $\mu$m peak-up photometry from Carpenter et al. (2006), and MIPS 24 \& 70 $\mu$m
fluxes or 3$\sigma$ upper limits from Carpenter et al. (2009, in preparation). The NextGen stellar atmospheric
models of Hauschildt et al. (1999) for the effective temperature nearest that of the assigned spectral type
are superposed, normalized using the $V-$band fluxes for the early-type stars and the $J-$band fluxes
for the late-type stars where photospheric emission peaks. The SED for the M0-type star J161420.2-190648
is normalized at $I-$band given its sharply rising near-infrared ($JHK_{S}$) flux levels.
\label{f1a}}
\end{figure}
\clearpage

\addtocounter{figure}{-1}
\addtocounter{subfigure}{1}
\clearpage
\begin{figure}
\epsscale{0.5}
\hspace{2cm}  \vspace{2cm}  \includegraphics[width=11cm,angle=0]{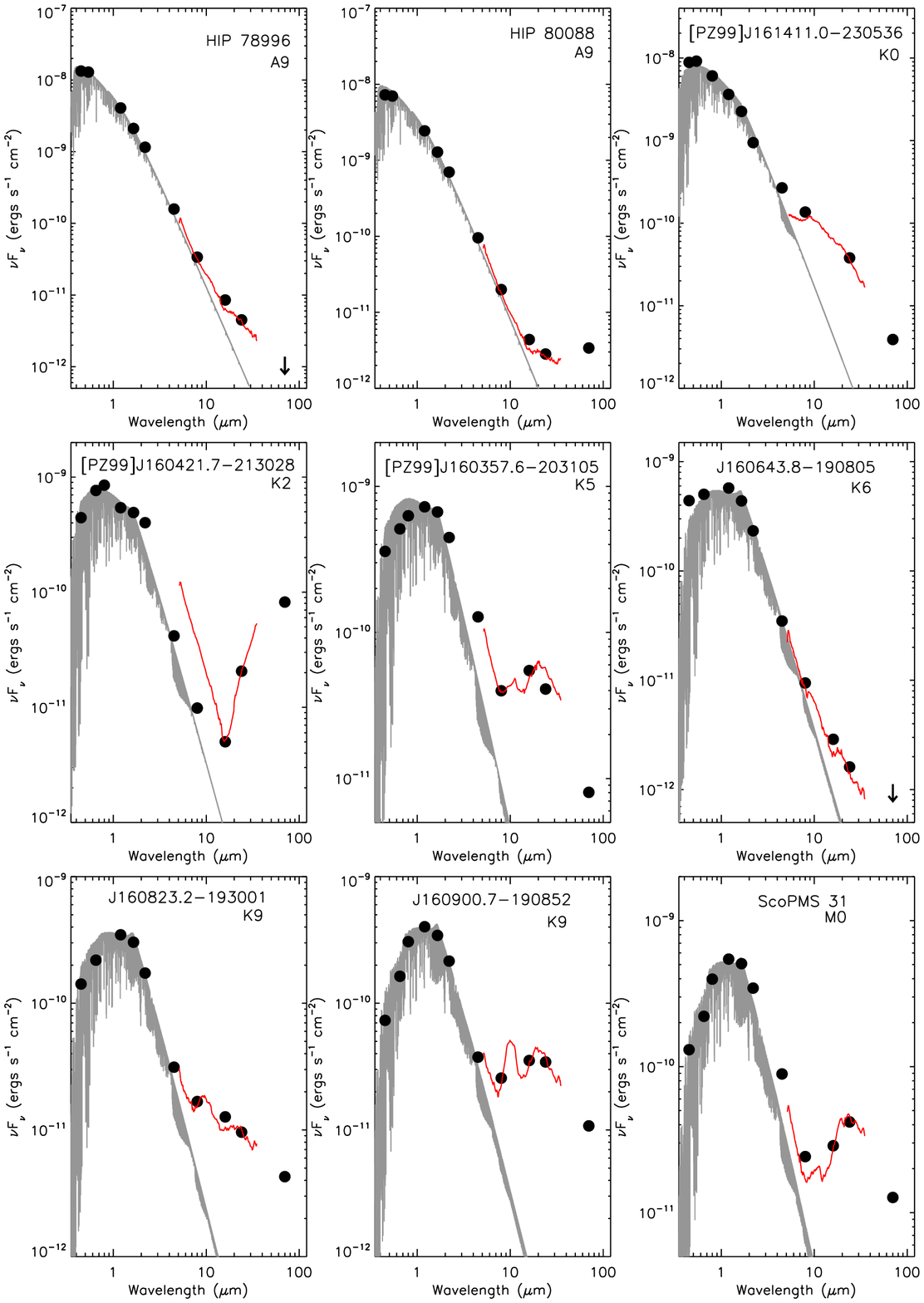}  \vspace{-2cm}    %  For Mac
\caption[f1b.ps]{b) Spectral energy distributions for the disk-bearing members of the Upper Scorpius OB association (continued).
\label{f1ab}}
\end{figure}
\clearpage

\addtocounter{figure}{-1}
\addtocounter{subfigure}{1}
\clearpage
\begin{figure}
\epsscale{0.5}
\hspace{2cm}  \vspace{2cm}  \includegraphics[width=11cm,angle=0]{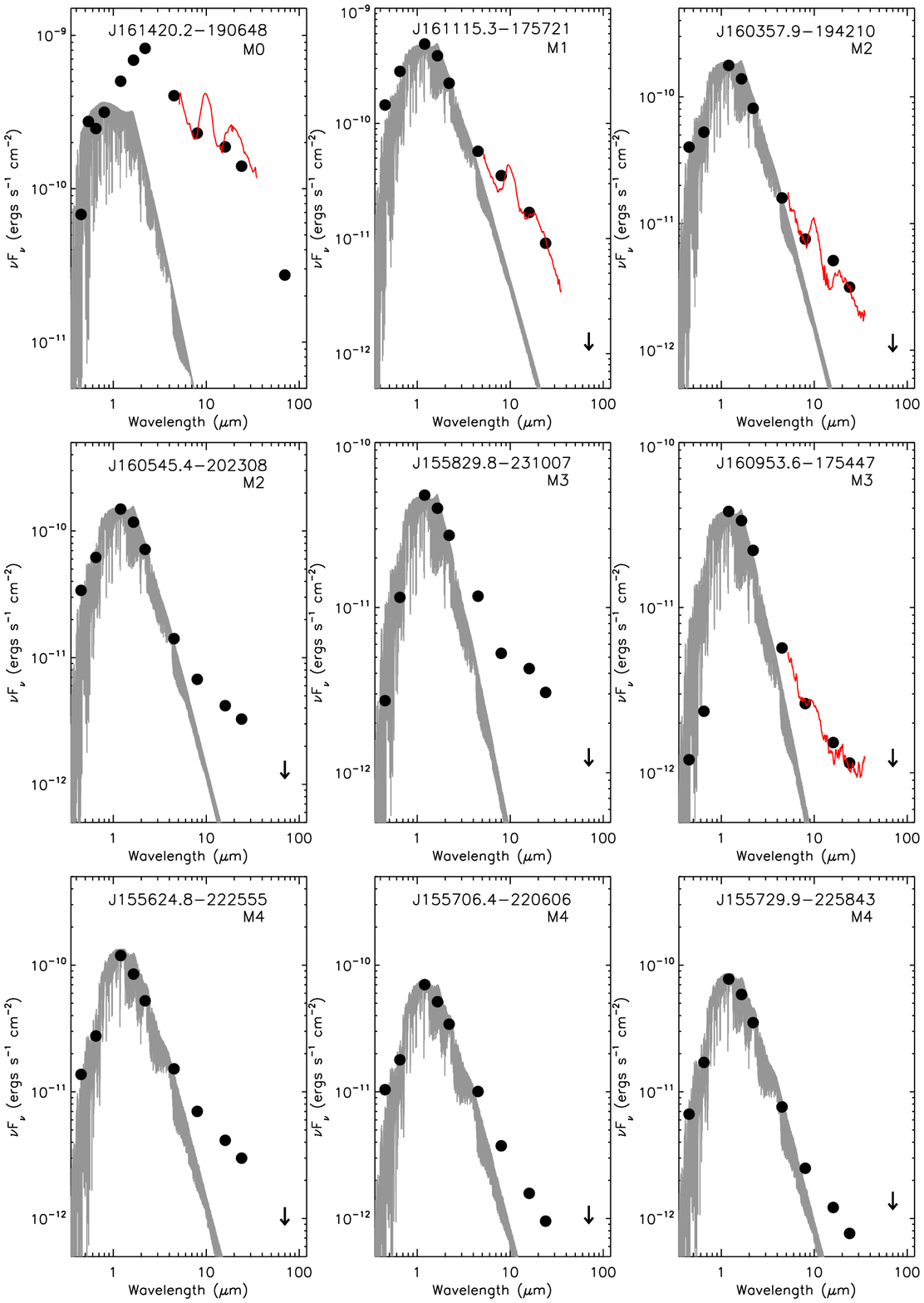}  \vspace{-2cm}    %  For Mac
\caption[f1c.ps]{c) Spectral energy distributions for the disk-bearing members of the Upper Scorpius OB association (continued).
\label{f1c}}
\end{figure}
\clearpage

\addtocounter{figure}{-1}
\addtocounter{subfigure}{1}
\clearpage
\begin{figure}
\epsscale{0.5}
\hspace{2cm}  \vspace{2cm}  \includegraphics[width=11cm,angle=0]{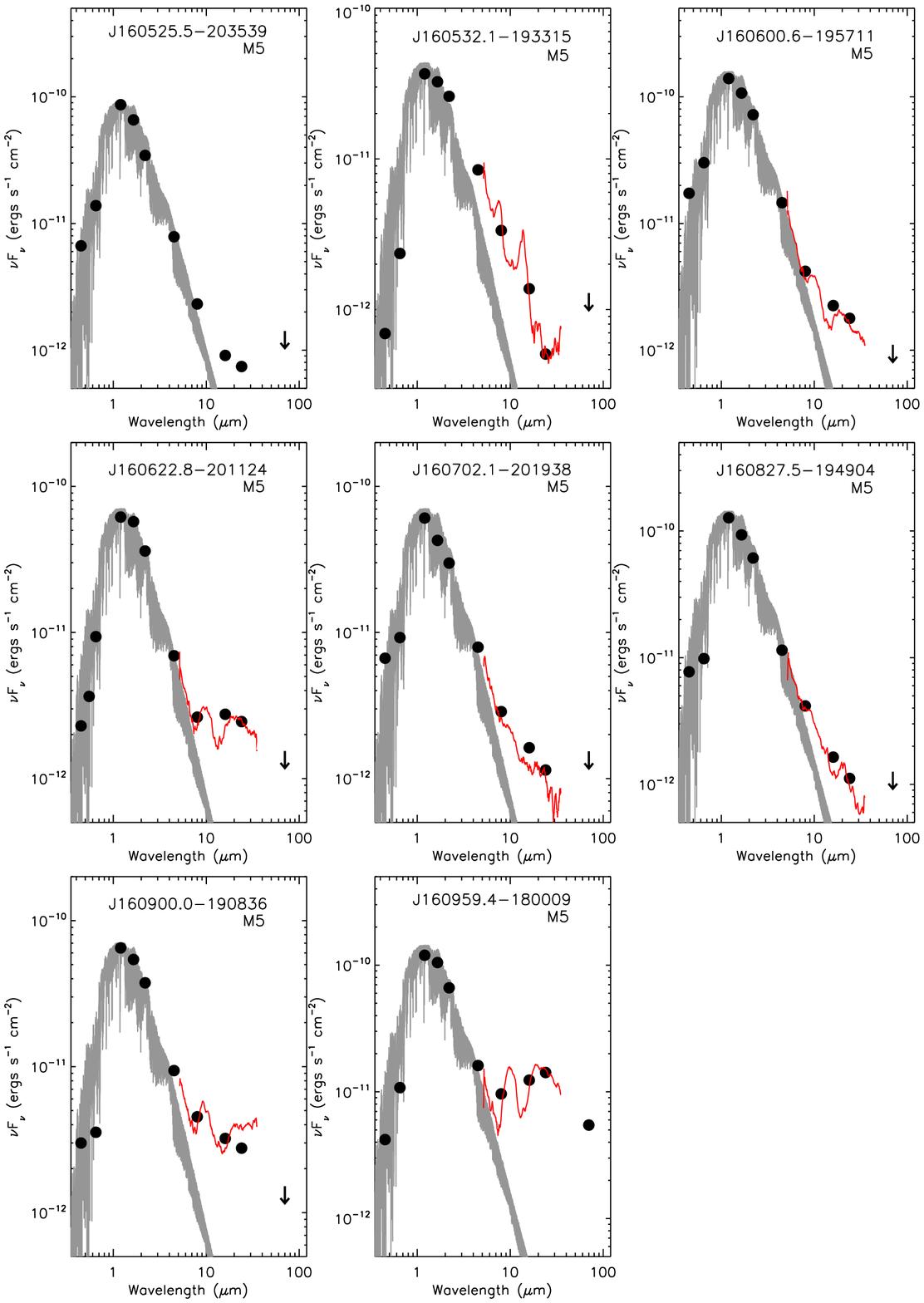}  \vspace{-2cm}    %  For Mac
\caption[f1d.ps]{d) Spectral energy distributions for the disk-bearing members of the Upper Scorpius OB association (continued).
\label{f1d}}
\end{figure}
\clearpage

\clearpage
\begin{figure}
\epsscale{0.5}
\hspace{2cm}  \vspace{2cm}  \includegraphics[width=11cm,angle=90]{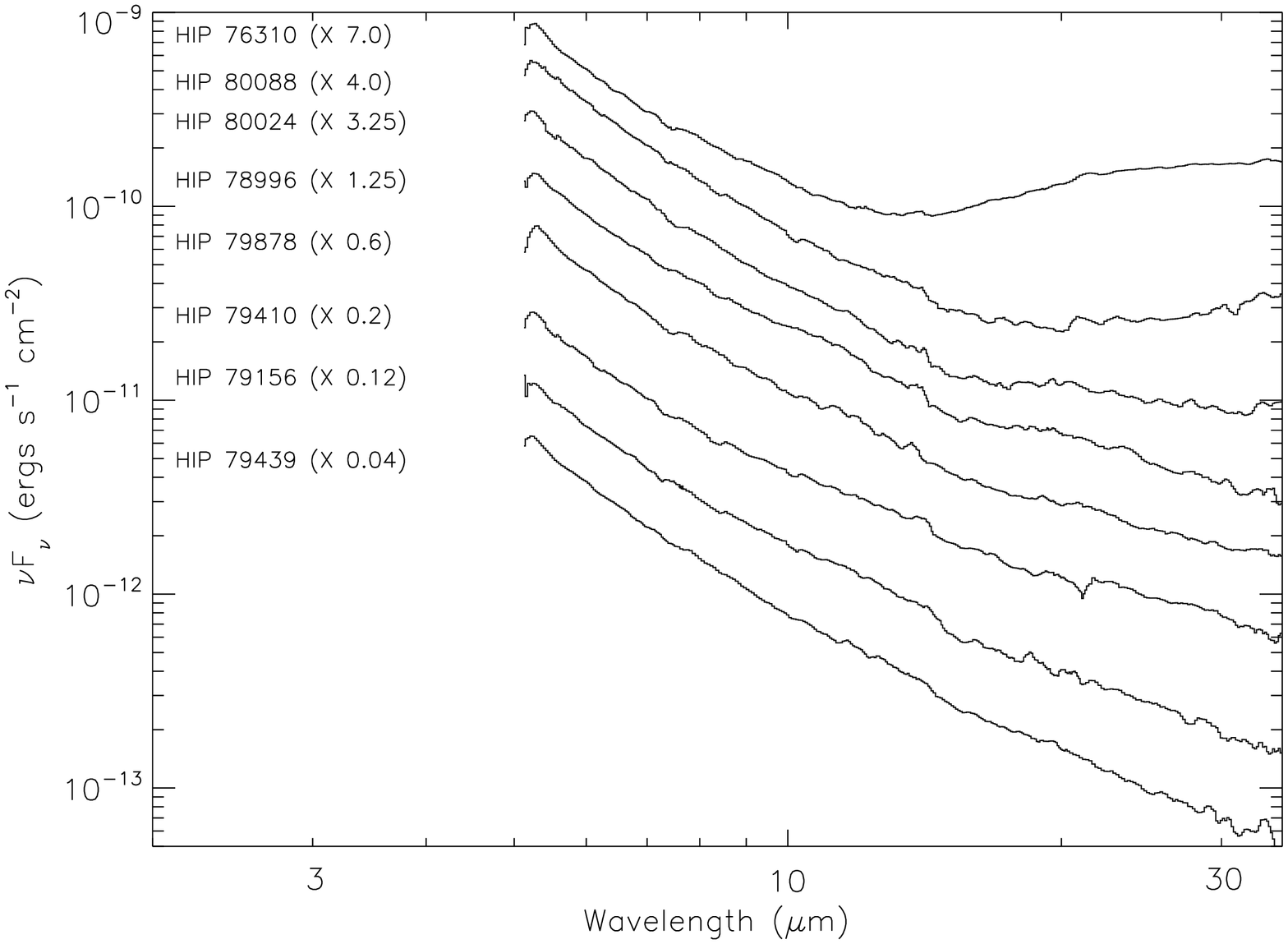}  \vspace{-2cm}    %  For Mac
\caption[f2.ps]{Morphological sequence of IRS mid-infrared spectra for early-type (B+A) Upper Scorpius 
excess members, arranged by SED shape and in approximate order of decreasing slope.
\label{f2}}
\end{figure}
\clearpage

\clearpage
\begin{figure}
\epsscale{0.5}
\hspace{2cm}  \vspace{2cm}  \includegraphics[width=11cm,angle=90]{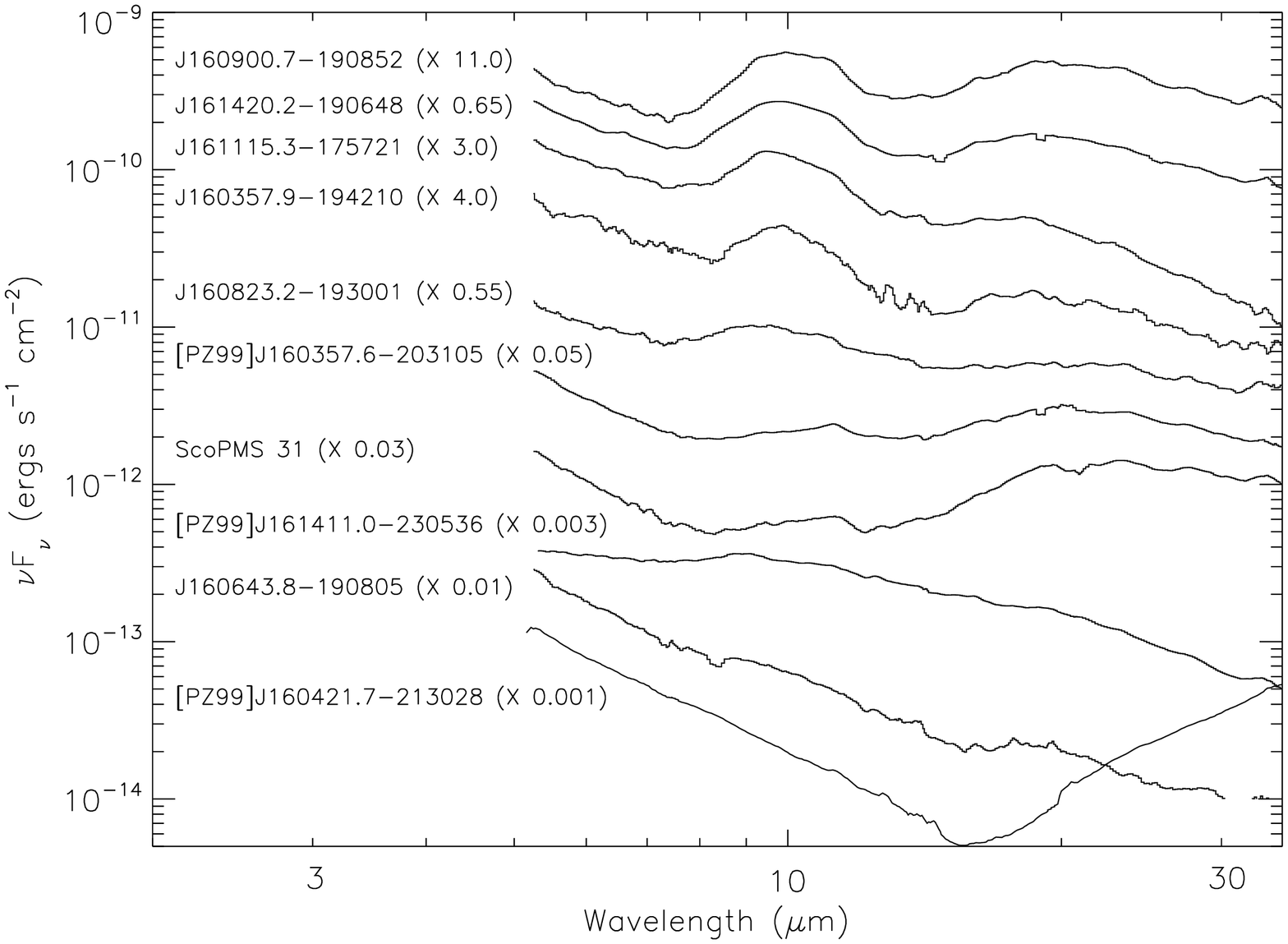}  \vspace{-2cm}    %  For Mac
\caption[f3.ps]{Morphological sequence of mid-infrared spectra for near solar-mass (K0--M2) 
Upper Scorpius excess members, arranged by SED shape and in approximate order of decreasing strength 
of the 10 $\mu$m silicate emission feature. The spectrum of the K2-type star [PZ99]J160421.7-213028, 
which lacks silicate emission features, is plotted at the bottom of the sequence.
\label{f3}}
\end{figure}
\clearpage

\clearpage
\begin{figure}
\epsscale{0.5}
\hspace{2cm}  \vspace{2cm}  \includegraphics[width=11cm,angle=90]{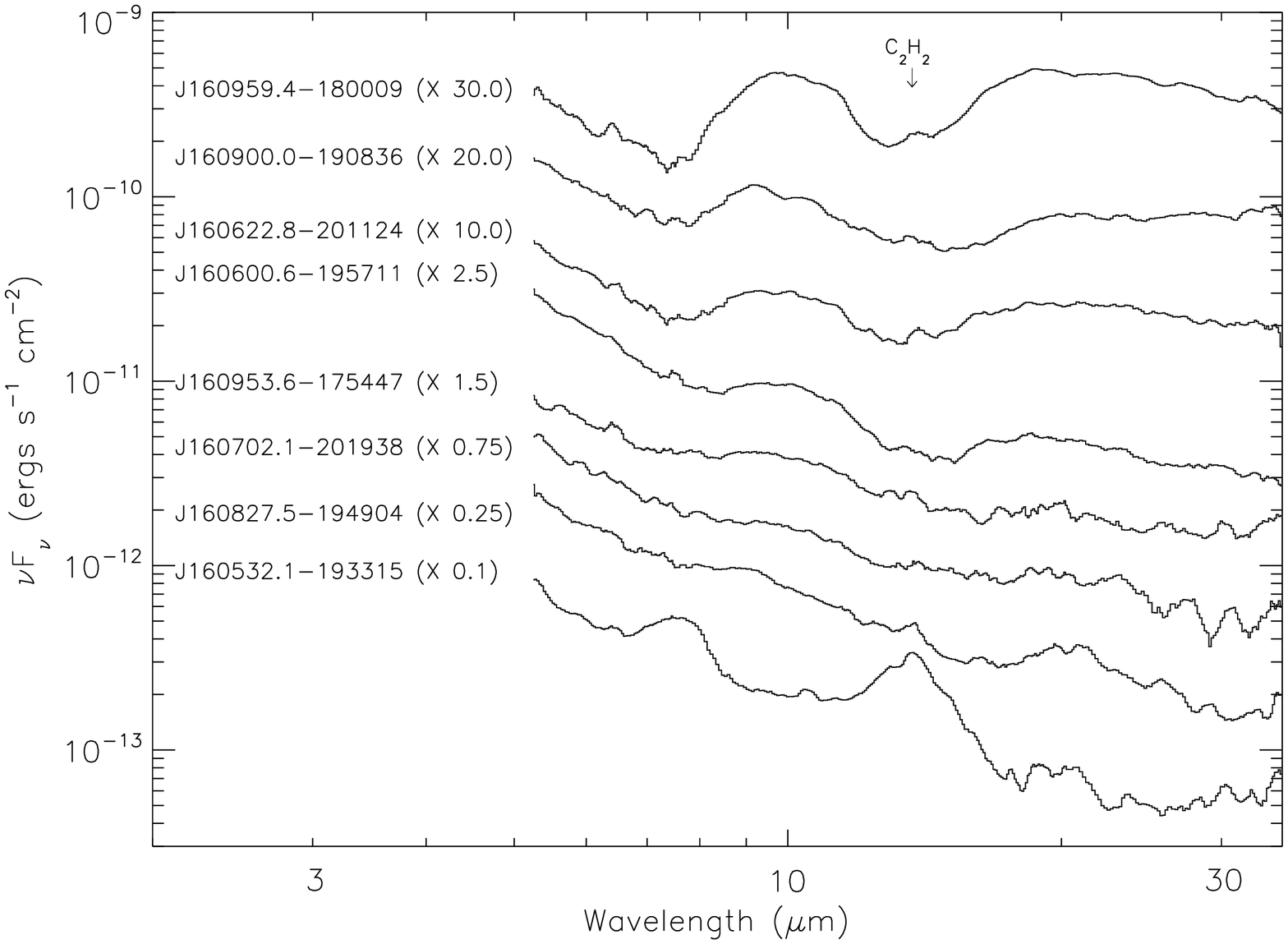}  \vspace{-2cm}    %  For Mac
\caption[f4.ps]{Morphological sequence of mid-infrared spectra for late-type (M3--M5) Upper Scorpius 
excess members, arranged by SED shape and decreasing strength of the 10 $\mu$m silicate emission feature.
The M5-type star J160532.1-193315 exhibits broad PAH emission near 7.7 $\mu$m and a strong emission
feature near 13.7 $\mu$m tentatively identified as the vibration-rotation band of C$_{2}$H$_{2}$. 
\label{f4}}
\end{figure}
\clearpage

\clearpage
\begin{figure}
\epsscale{0.5}
\hspace{2cm}  \vspace{2cm}  \includegraphics[width=11cm,angle=90]{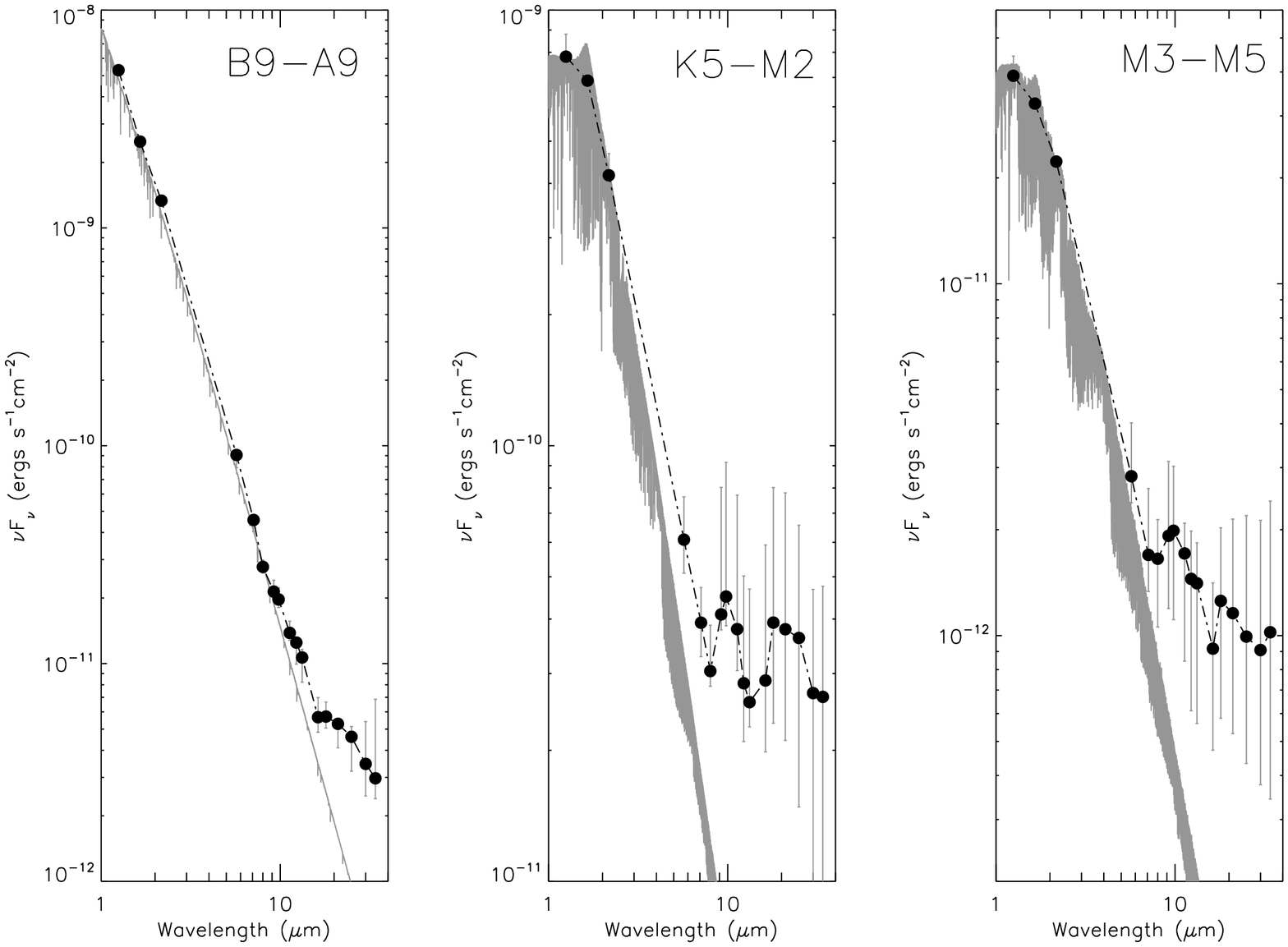}  \vspace{-2cm}    %  For Mac
\caption[f5.ps]{The median SED from 1.25 to 34.0 $\mu$m for early-type (B9--A9) stars (left),
near solar-mass (K5--M2) stars (middle), and low-mass (M3--M5) stars (right). The error bars
define the upper and lower quartiles of the observed fluxes. Superposed are the NextGen
stellar atmospheric models of Hauschildt et al. (1999), fitted near $H-$band, for $T_{\rm eff}$
of 9600 K (A0), 3800 K (M0), and 3200 K (M5), respectively.
\label{f5}}
\end{figure}
\clearpage

\clearpage
\begin{figure}
\epsscale{0.5}
\hspace{2cm}  \vspace{2cm}  \includegraphics[width=11cm,angle=0]{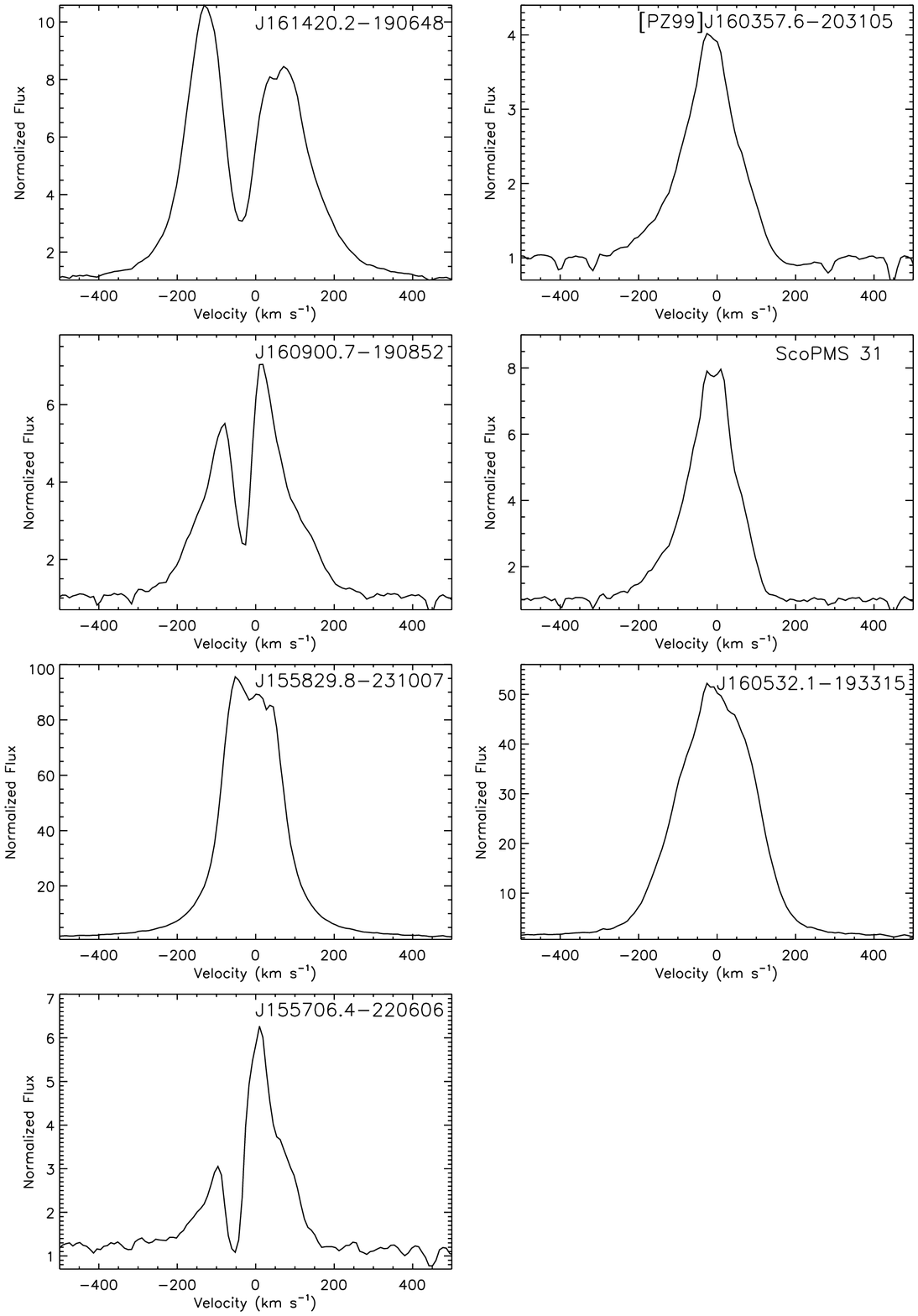}  \vspace{-2cm}    %  For Mac
\caption[f6.ps]{H$\alpha$ emission line profiles for the 7 suspected accretors among the 35 Upper Scorpius
excess sources observed with HIRES. All exhibit velocity widths at 10\% peak flux of $>$ 270 km s$^{-1}$,
an indicator of accretion using the criterion established by White \& Basri (2003) for optically veiled
pre-main sequence stars.
\label{f6}}
\end{figure}
\clearpage

\clearpage
\begin{figure}
\epsscale{0.5}
\hspace{2cm}  \vspace{2cm}  \includegraphics[width=11cm,angle=0]{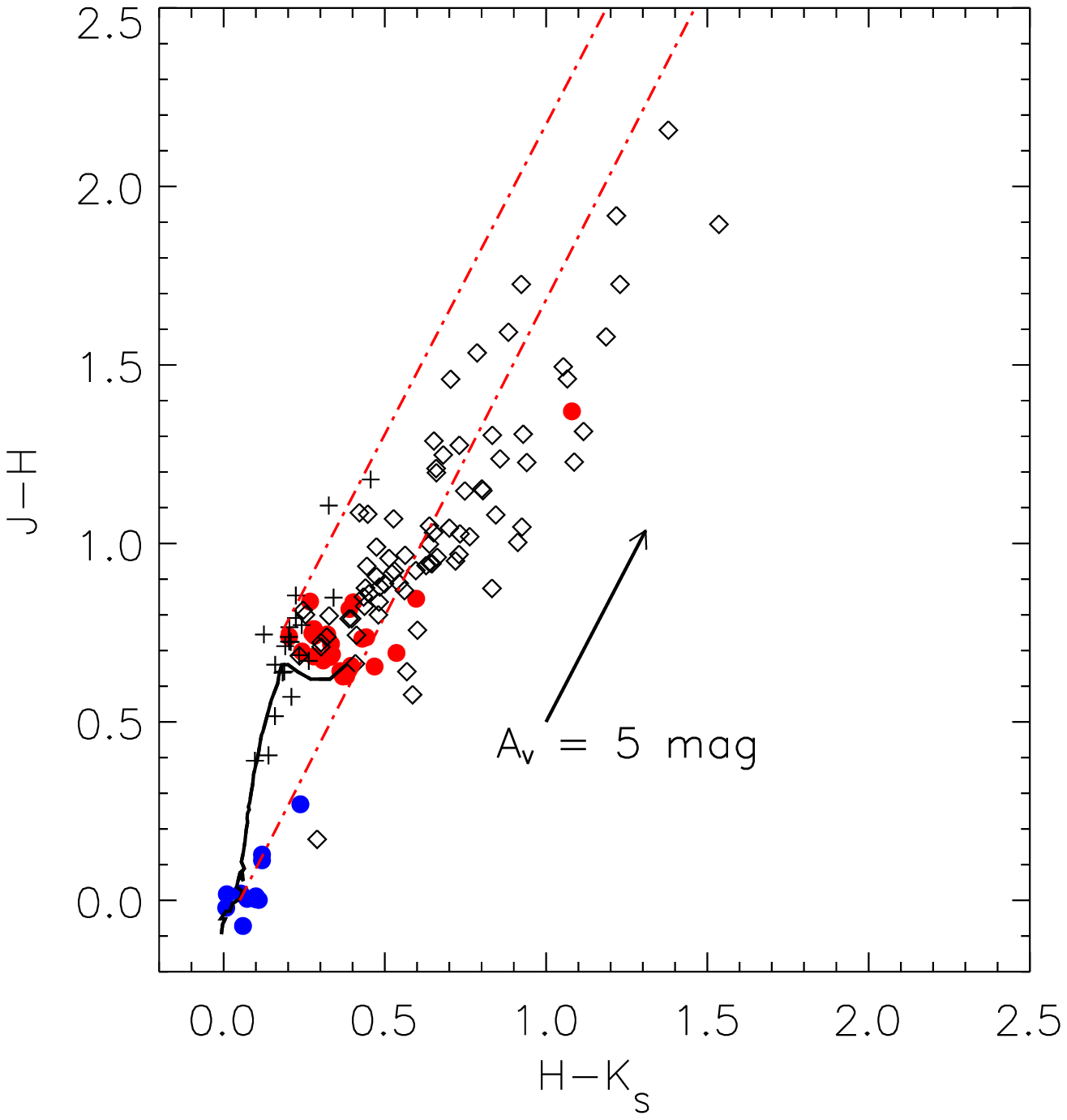}  \vspace{-2cm}    %  For Mac
\caption[f7.ps]{The 2MASS $H-K_{S}$, $J-H$ color-color diagram for the Upper Scorpius infrared excess stars,
71 Class II sources (open diamonds), and 17 Class III sources (plus symbols) of the Taurus-Auriga star 
forming region from Furlan et al. (2006). The 35 Upper Scorpius members appear clustered around the 
locus of main sequence stars, with the early-type members (blue) lying near the base of the trunk and 
the late-type stars (red) just above the M-dwarf branch. The Taurus-Auriga Class II sources suffer 
significant extinction and a large fraction (over half) exhibit colors that place them outside of the 
reddening boundaries for normal dwarfs. 
\label{f7}}
\end{figure}
\clearpage

\clearpage
\begin{figure}
\epsscale{0.5}
\hspace{2cm}  \vspace{2cm}  \includegraphics[width=11cm,angle=0]{f8}  \vspace{-2cm}    %  For Mac
\caption[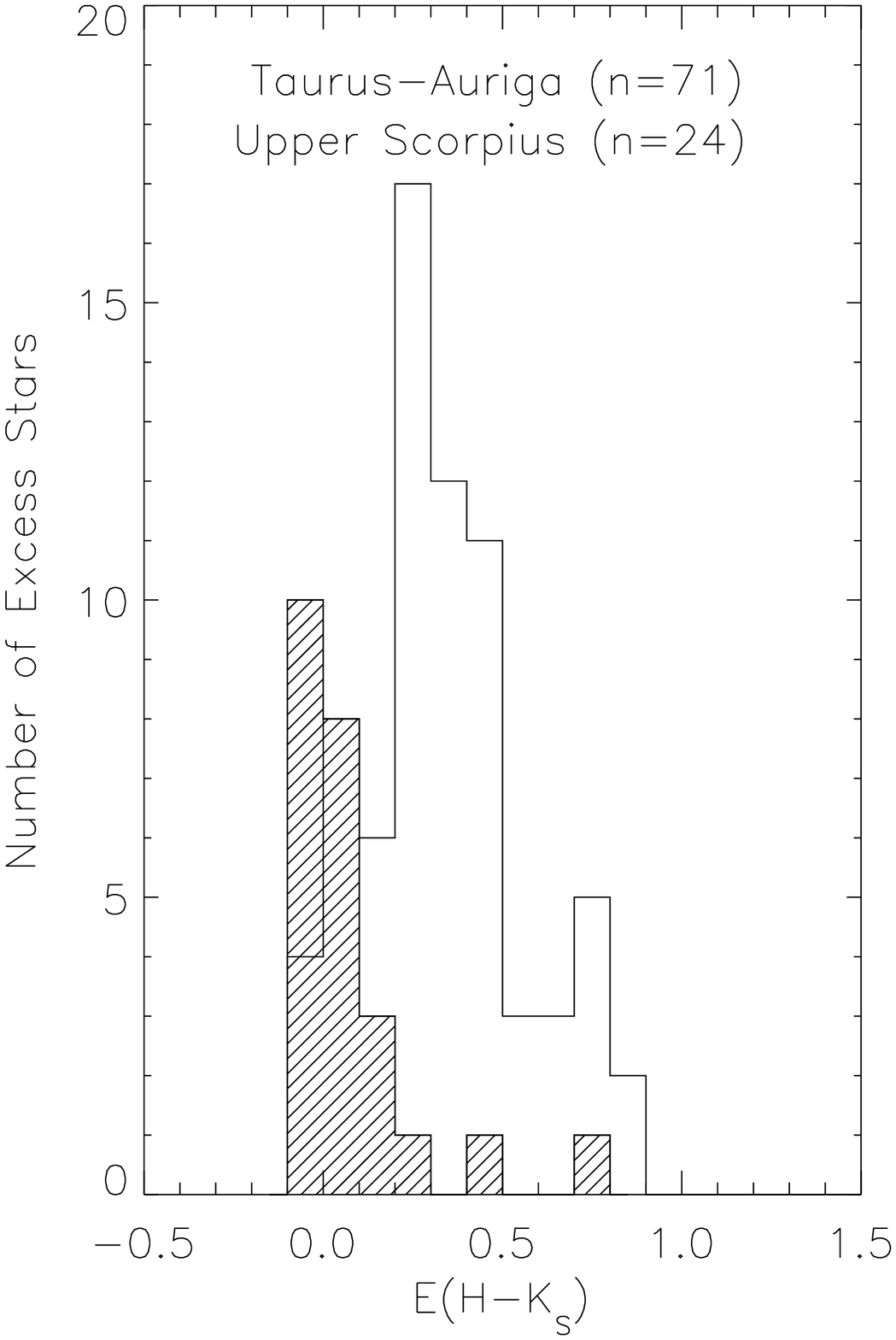]{The $E(H-K_{S})$ histograms for Upper Scorpius excess members (cross-hatched region) and 
Taurus-Auriga Class II sources (open region). Intrinsic $H-K_{S}$ colors for B2--K5 type stars were 
obtained from a tabulation of main sequence 2MASS colors adopted by the FEPS survey (Carpenter et al. 2008).
For later spectral types, intrinsic 2MASS $H-K_{S}$ colors were derived using $\sim$1100 
stars from the Palomar/MSU Nearby Star Spectroscopic Survey of Reid et al. (1995) and Hawley et al. (1996).
The distributions of color-excesses are markedly different, with median $E(H-K_{S})$ values of 0.03 and
0.35 for the Upper Scorpius and Taurus-Auriga samples, respectively.
\label{f8}}
\end{figure}
\clearpage

\clearpage
\begin{figure}
\epsscale{0.5}
\hspace{2cm}  \vspace{2cm}  \includegraphics[width=11cm,angle=0]{f9}  \vspace{-2cm}    %  For Mac
\caption[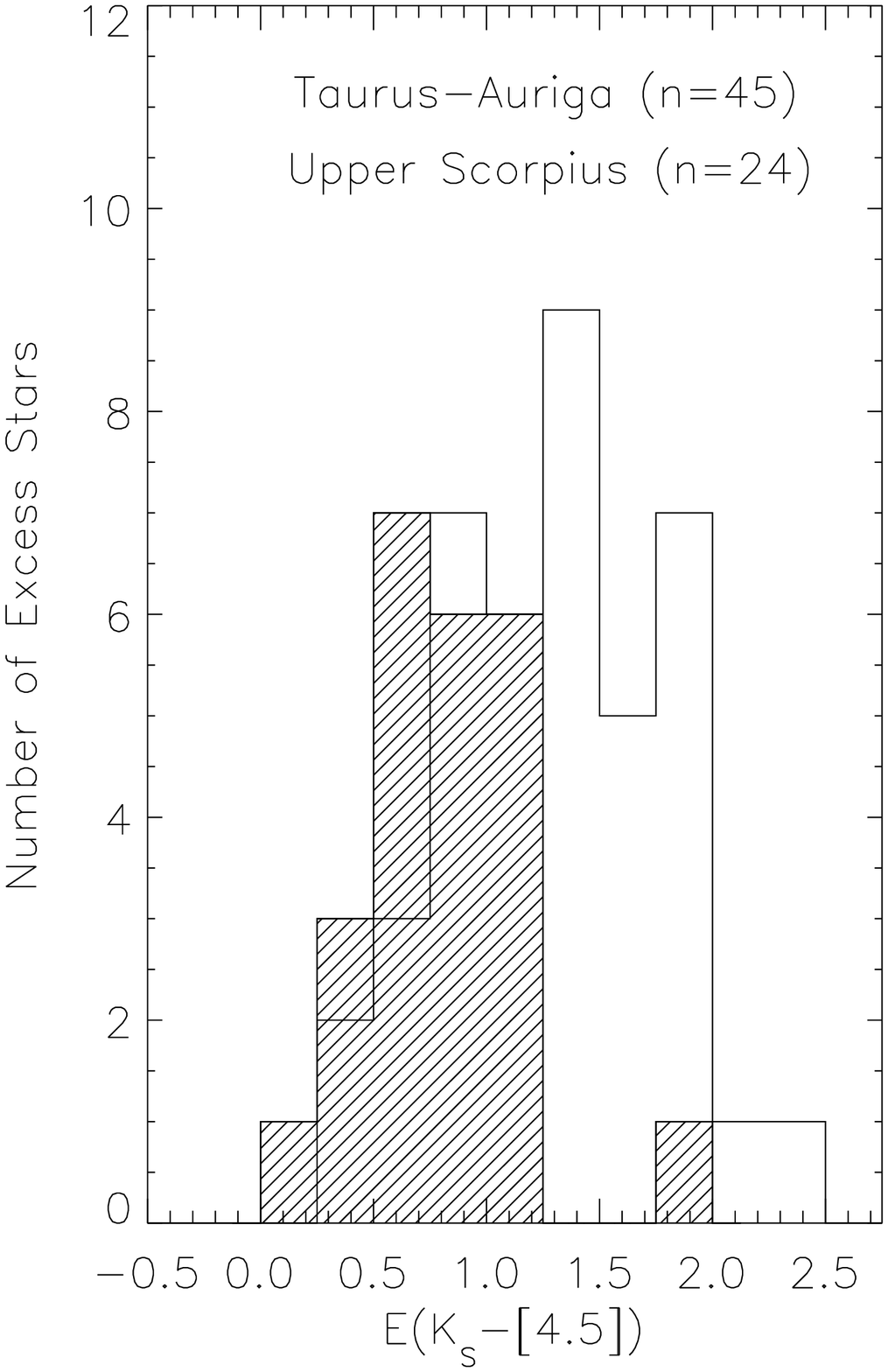]{The $E(K_{S}-[4.5])$ histograms for the 24 Upper Scorpius low-mass excess members (cross-hatched region) 
and 45 Taurus-Auriga Class II sources (open region) with IRAC photometry available from Hartmann et al. (2005), 
Luhman et al. (2007), and the Taurus Legacy program of Padgett et al. (2008). Intrinsic $K_{S}-[4.5]$ 
colors were derived using extinction corrected 2MASS $K_{S}$ magnitudes and stellar photospheric 4.5 $\mu$m 
fluxes estimated using the {\it Spitzer} Stellar Performance Estimation Tool (STAR-PET). The $E(K_{S}-[4.5])$
distributions are markedly different with median color excesses of 1.37 and 0.77 mag for the Taurus-Auriga and 
Upper Scorpius samples, respectively.
\label{f9}}
\end{figure}
\clearpage

\clearpage
\begin{figure}
\epsscale{0.5}
\hspace{2cm}  \vspace{2cm}  \includegraphics[width=11cm,angle=0]{f10}  \vspace{-2cm}    %  For Mac
\caption[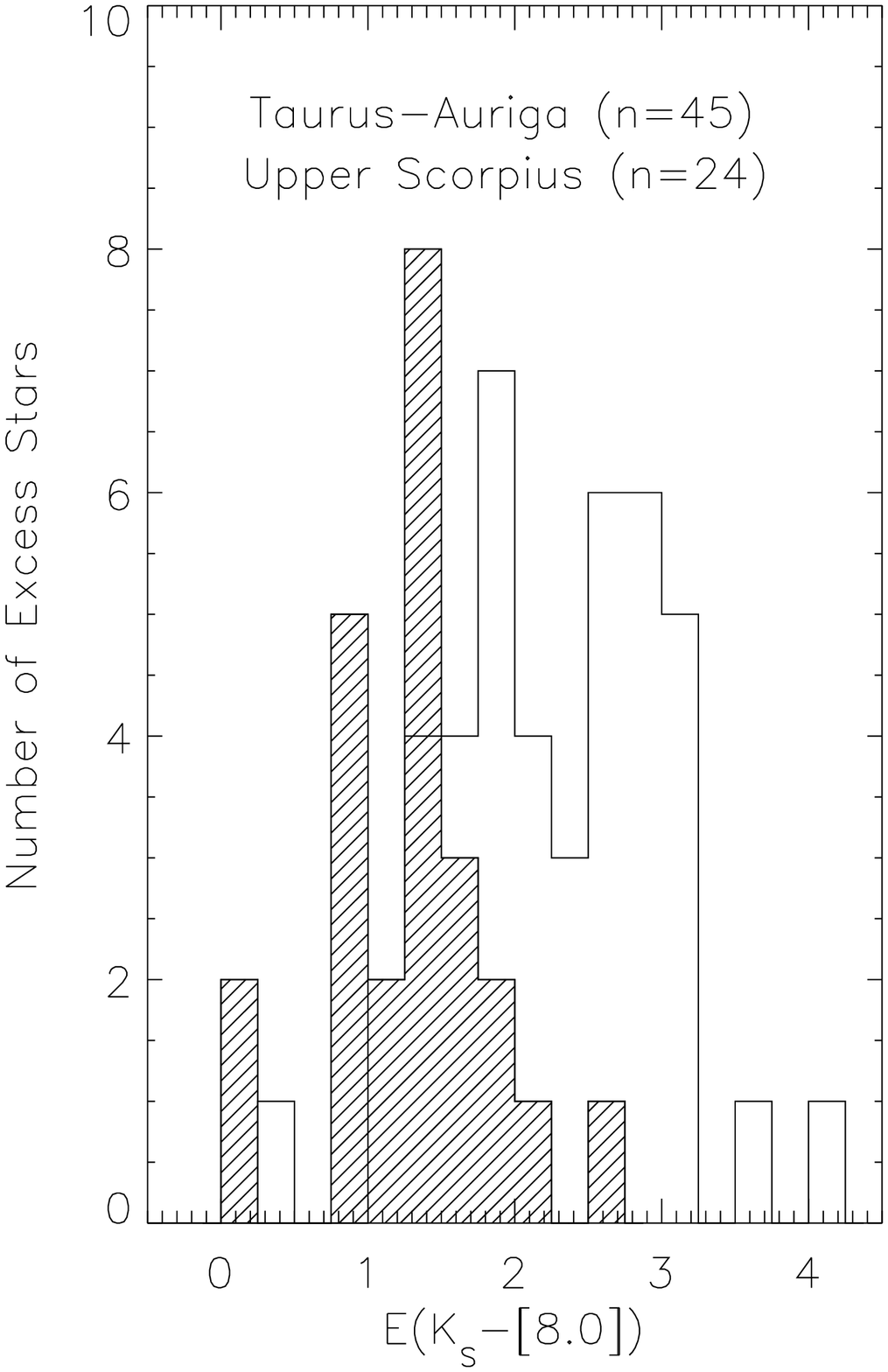]{The $E(K_{S}-[8.0])$ histograms for the 24 Upper Scorpius low-mass excess members (cross-hatched region)
and 45 Taurus-Auriga Class II sources (open region) with IRAC photometry available from Hartmann et al. (2005), 
Luhman et al. (2007), or the Taurus Legacy program of Padgett et al. (2008). Intrinsic $K_{S}-[8.0]$
colors were derived using extinction corrected 2MASS $K_{S}$ magnitudes and stellar photospheric 8.0 $\mu$m 
fluxes estimated using the {\it Spitzer} Stellar Performance Estimation Tool (STAR-PET). The $E(K_{S}-[8.0])$
distributions are markedly different with median values of 2.23 and 1.43 mag for Taurus-Auriga and Upper Scorpius, 
respectively. Thermal and mid-infrared excesses are significantly different among these populations suggesting
significant differences in disk structure, possibly the result of evolution.
\label{f10}}
\end{figure}
\clearpage

\clearpage
\begin{figure}
\epsscale{0.5}
\hspace{2cm}  \vspace{2cm}  \includegraphics[width=11cm,angle=0]{f11}  \vspace{-2cm}    %  For Mac
\caption[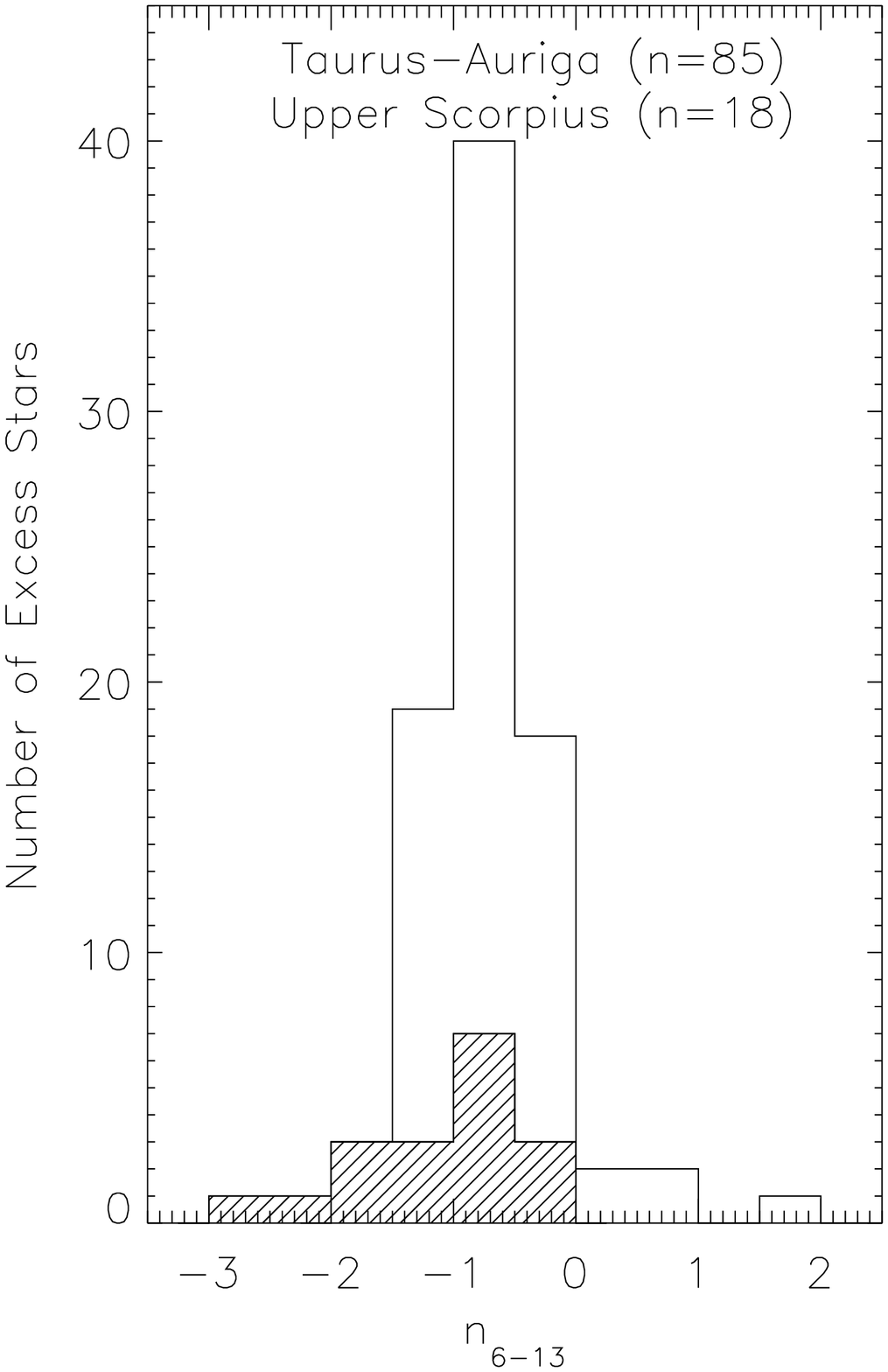]{Histogram of $n_{6-13}$ spectral indices for the 18 Upper Scorpius low-mass excess members (cross-hatched) observed
with IRS and the 85 Class II sources in Taurus-Auriga (open) from Furlan et al. (2006). For accreting primordial disks, the $n_{6-13}$
spectral index is predominantly impacted by the observed inclination angle and less so by the mass accretion rate.
\label{f11}}
\end{figure}
\clearpage

\clearpage
\begin{figure}
\epsscale{0.5}
\hspace{2cm}  \vspace{2cm}  \includegraphics[width=11cm,angle=0]{f12}  \vspace{-2cm}    %  For Mac
\caption[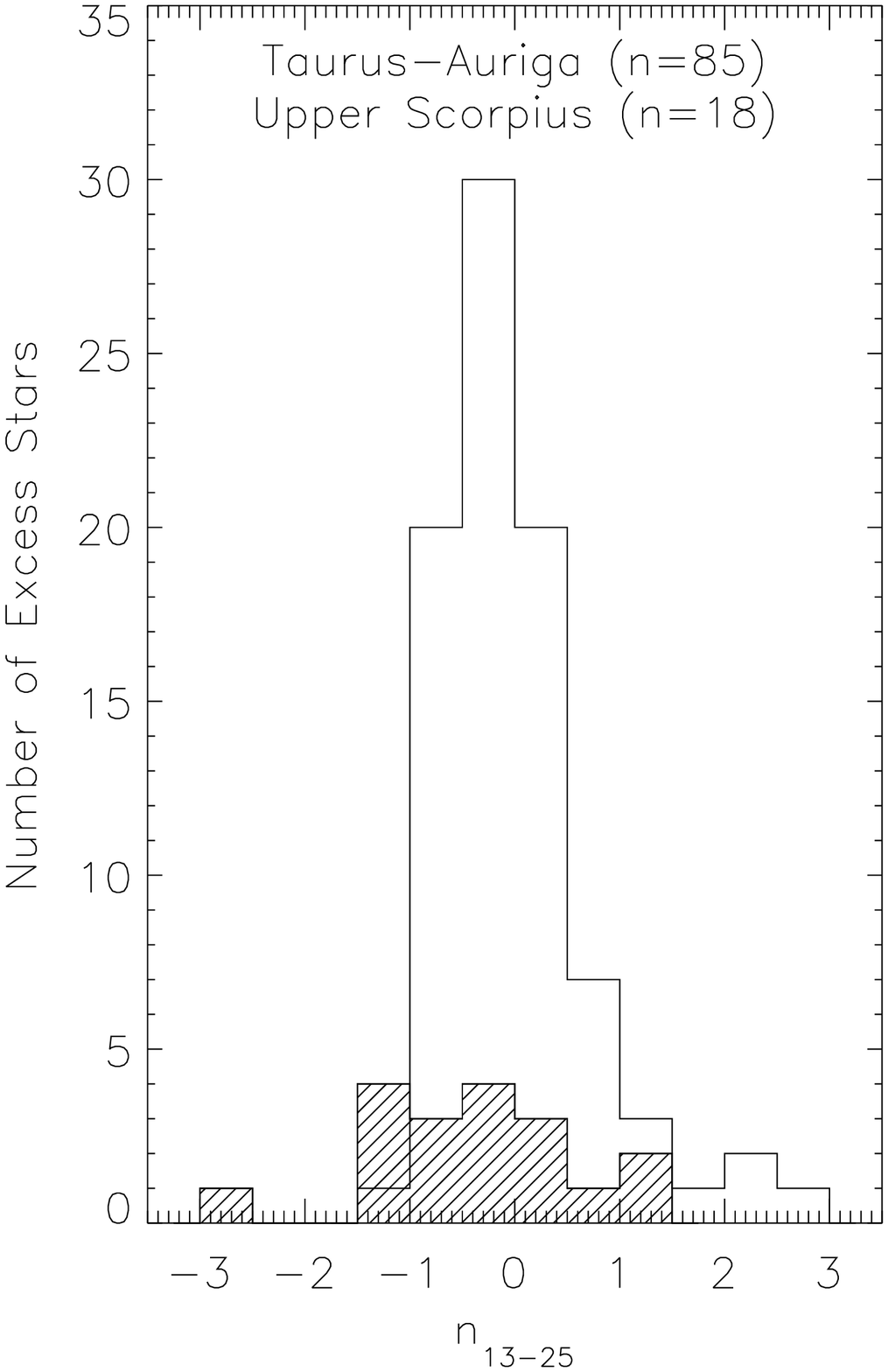]{Histogram of $n_{13-25}$ spectral indices for the 18 Upper Scorpius low-mass excess members (cross-hatched) observed
with IRS and the 85 Class II sources in Taurus-Auriga (open) from Furlan et al. (2006). The $n_{13-25}$ spectral index is predominantly
impacted by the amount of dust grain growth and mid-plane settling within the disk.
\label{f12}}
\end{figure}
\clearpage

\clearpage
\begin{figure}
\epsscale{0.5}
\hspace{2cm}  \vspace{2cm}  \includegraphics[width=11cm,angle=90]{f13}  \vspace{-2cm}    %  For Mac
\caption[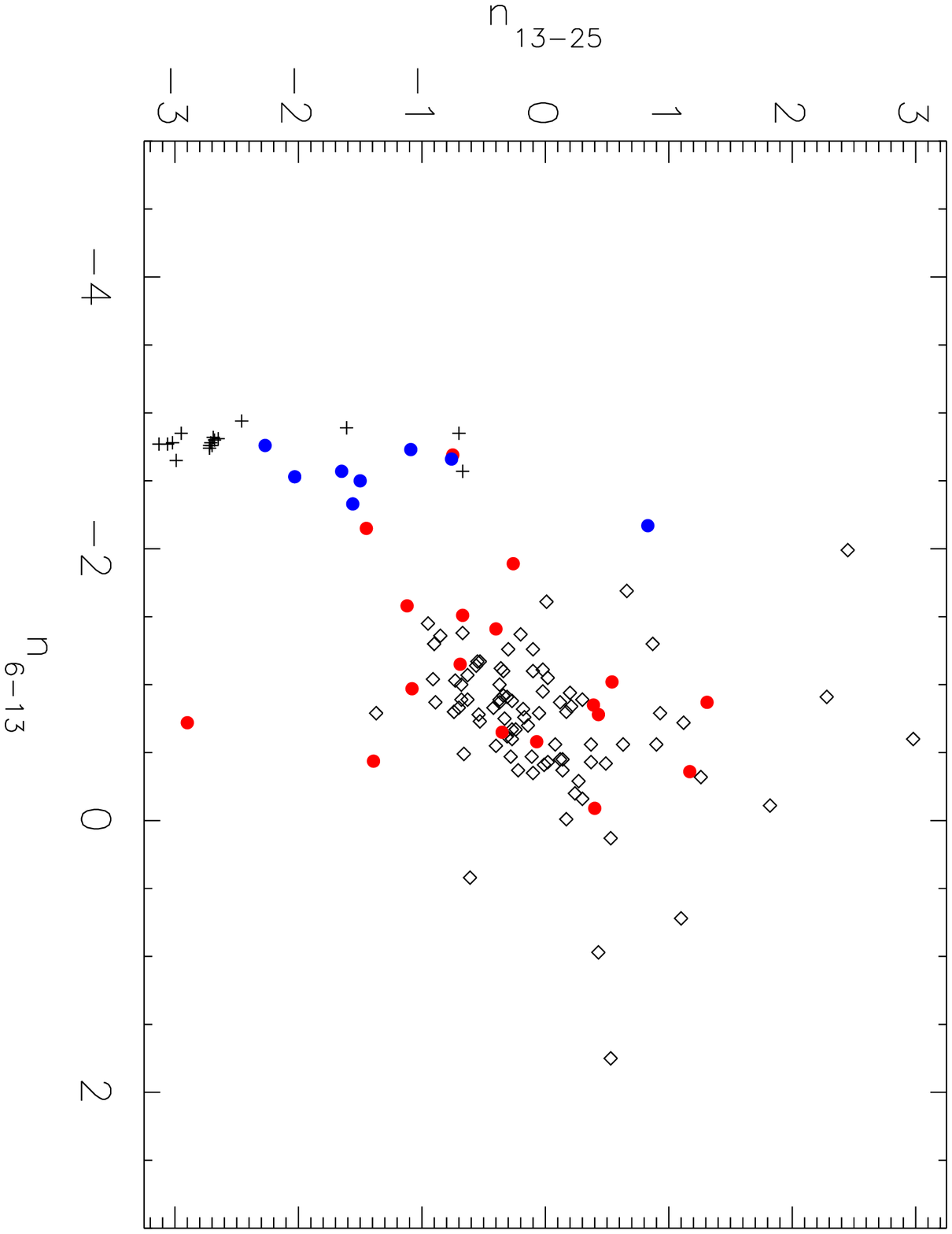]{The $n_{13-25}$ mid-infrared spectral index plotted against the $n_{6-13}$ index for the 26 excess members
of Upper Scorpius observed with IRS and the 85 Class II and 26 Class III sources in Taurus-Auriga from Furlan et al. (2006).
Symbols are as in Figure 9. The Upper Scorpius member near $n_{6-13}$$\sim$$-$0.75, $n_{13-25}$$\sim$$-$2.8 is the M5-type
star, J160532.1-193315, which exhibits strong emission features near 7.7 and 13.7 $\mu$m, attributed to PAHs and the 
vibration-rotation band of C$_{2}$H$_{2}$, respectively.
\label{f13}}
\end{figure}
\clearpage

\clearpage
\begin{figure}
\epsscale{0.5}
\hspace{2cm}  \vspace{2cm}  \includegraphics[width=11cm,angle=90]{f14}  \vspace{-2cm}    %  For Mac
\caption[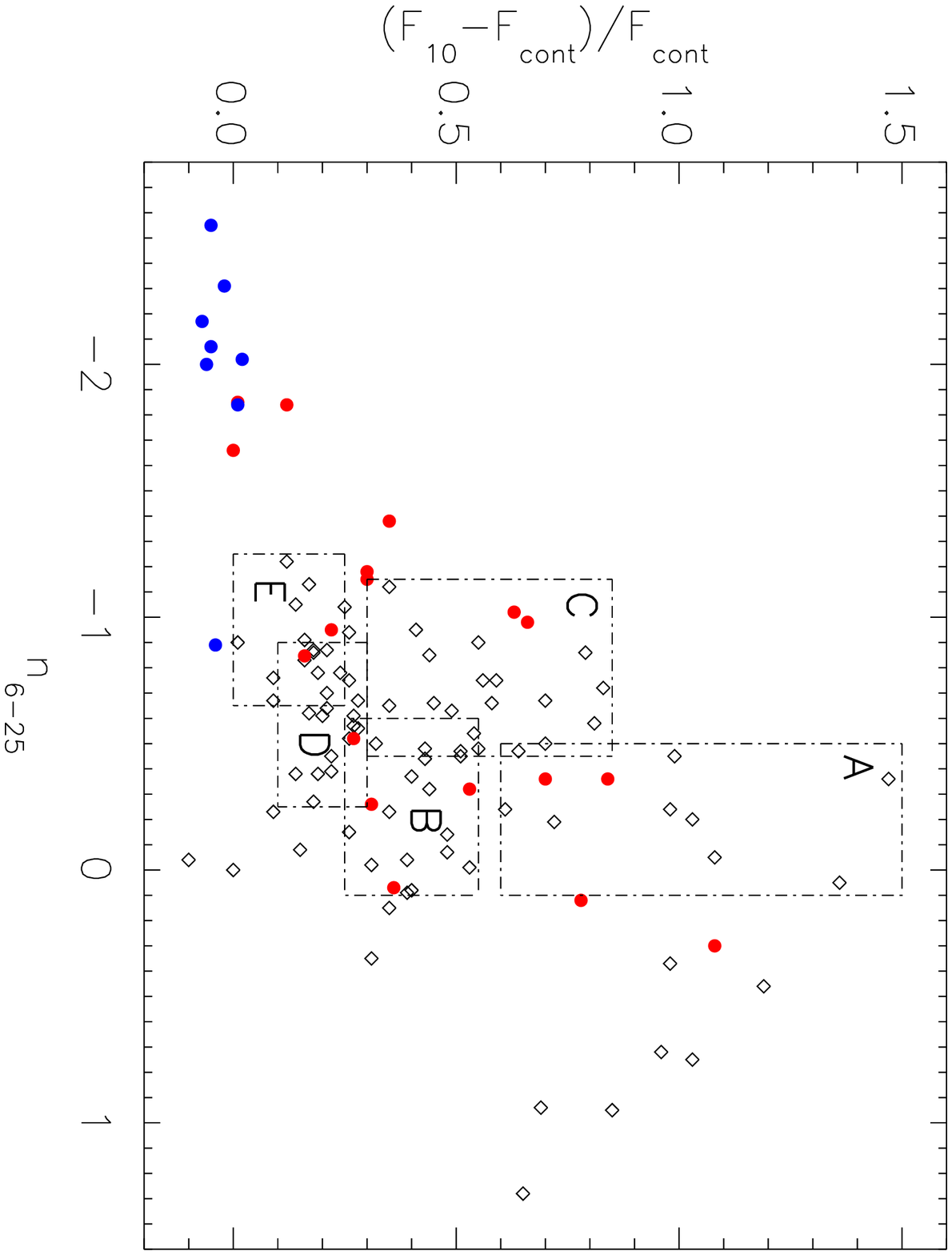]{The continuum-subtracted, integrated flux of the 10 $\mu$m silicate emission feature, normalized to
the continuum and plotted as a function of the $n_{6-25}$ spectral index for the 26 Upper Scorpius members 
observed with IRS and the 85 Class II sources in Taurus-Auriga from Furlan et al. (2006). Symbols are as in Figure 9. 
Also shown are the approximate boundaries of the various classes (A--E) in the morphological sequence of mid-infrared
SEDs defined by Furlan et al. (2006). In general, stars with $n_{6-25}$ $< -1.0$ exhibit weak 10 $\mu$m silicate 
emission. 
\label{f14}}
\end{figure}
\clearpage

\clearpage
\begin{figure}
\epsscale{0.5}
\hspace{2cm}  \vspace{2cm}  \includegraphics[width=11cm,angle=0]{f15}  \vspace{-2cm}    %  For Mac
\caption[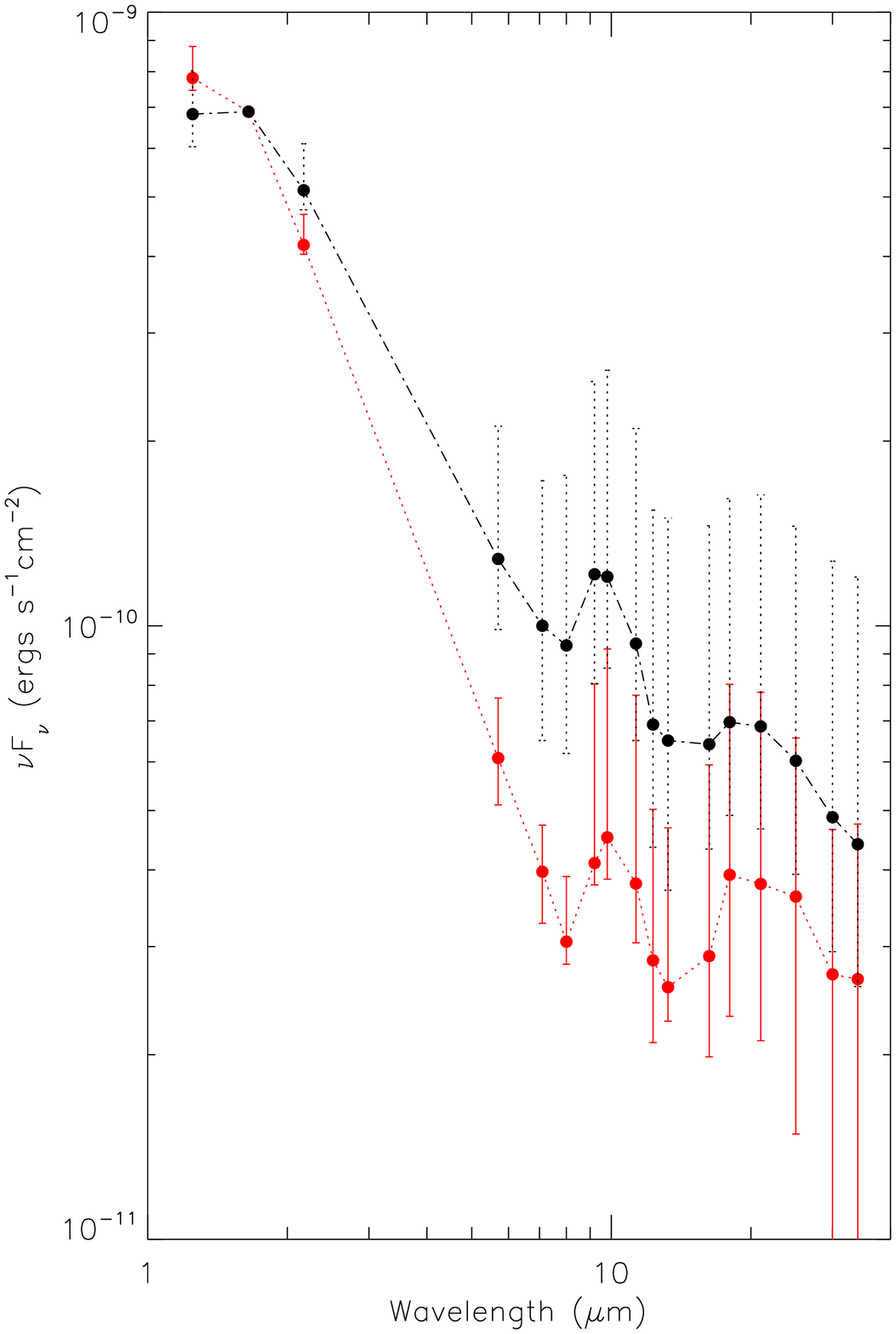]{The median SED from 1.25 to 34.0 $\mu$m derived using 2MASS $JHK_{S}$ photometry 
and IRS spectra for the 8 Upper Scorpius near solar-mass (K5--M2) excess members (red curve). The 
error bars define the upper and lower quartiles of the observed fluxes. Also shown (black curve)
is the median SED for the 55 K5--M2 type Class II sources in the Taurus-Auriga sample of Furlan et al. (2006). 
\label{f15}}
\end{figure}
\clearpage

\clearpage
\begin{figure}
\epsscale{0.5}
\hspace{2cm}  \vspace{2cm}  \includegraphics[width=11cm,angle=0]{f16}  \vspace{-2cm}    %  For Mac
\caption[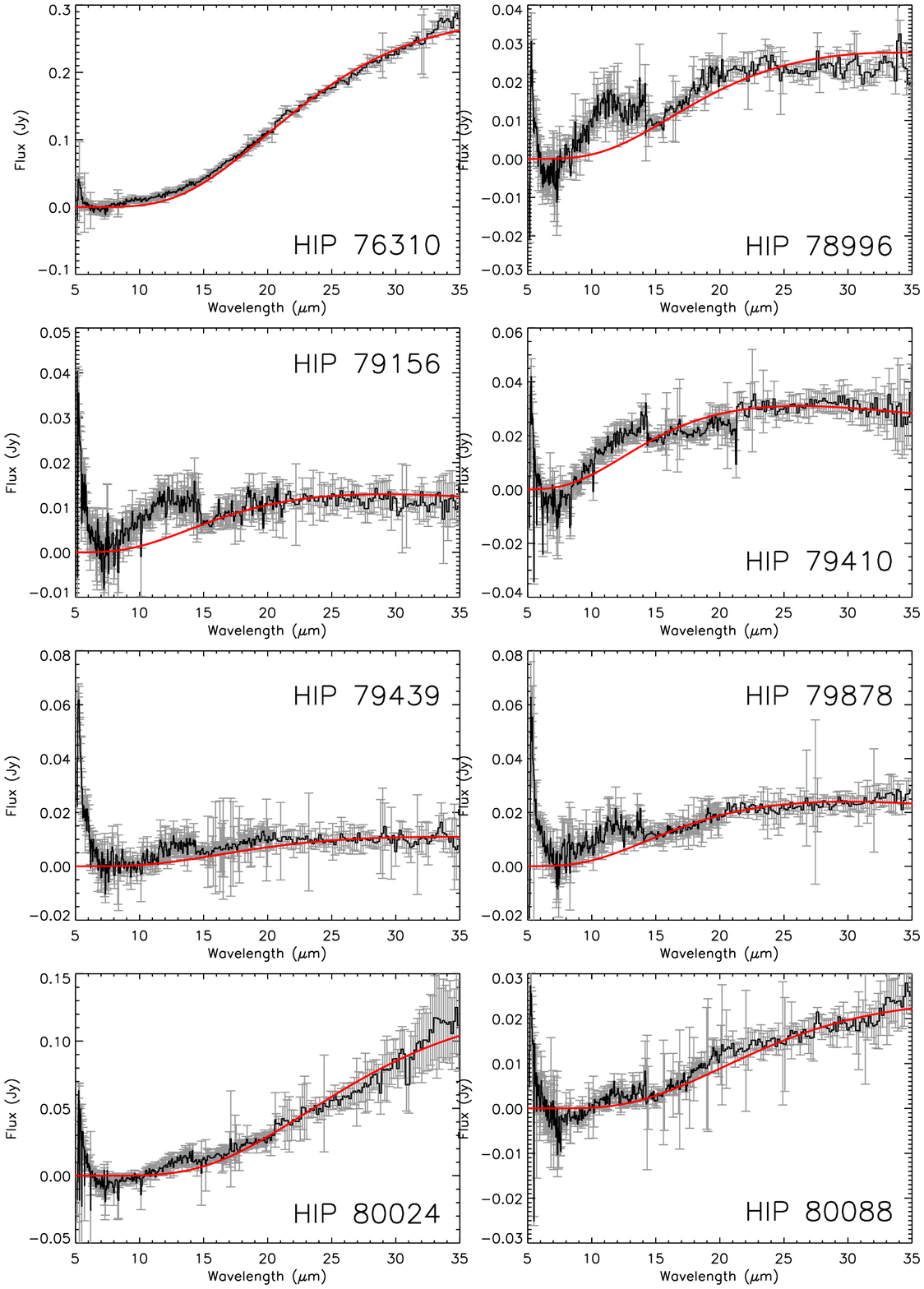]{Photosphere-subtracted IRS spectra of the early-type (B9--A9) debris disk members of
Upper Scorpius. Superposed in red are the single-temperature blackbody SEDs used to estimate the dust
grain temperatures. The slight rise in flux levels from 10--15 $\mu$m is possibly attributed to a 
discontinuity between the SL and LL modules. These features are ignored when fitting the single-temperature
blackbodies to the photosphere-subtracted SED.
\label{f16}}
\end{figure}
\clearpage

\begin{deluxetable}{cccccccccccccc}
\tabletypesize{\tiny}
\rotate
\tablenum{1}
\tablewidth{0pt}
\tablecaption{Observed Spectroscopic Properties of Upper Scorpius Excess Members}
\tablehead{
\colhead{Source}  & \colhead{ST\tablenotemark{a}}  & \colhead{$A_{V}$\tablenotemark{b}}  & \colhead{V$_{\rm r}$\tablenotemark{c}}  & \colhead{W(H$\beta$)\tablenotemark{d}} &  \colhead{W(He I)\tablenotemark{d}} &  \colhead{W(H$\alpha$)\tablenotemark{d}} & \colhead{W(Li I)} &  \colhead{W(Ca II)\tablenotemark{e}}  &   \colhead{Other Emission\tablenotemark{f}} &  \colhead{Comments\tablenotemark{g}} \\
       &     & (mag)  &  (km s$^{-1}$) &  (\AA)      &  $\lambda$5876(\AA) &  (\AA)          & $\lambda$6708(\AA)  &  $\lambda$8542(\AA) &                      &           \\
}
\startdata
HIP 77859              & B2 & 0.5 & $-$8.6           &  abs      &  abs    &  abs      & ...   &  abs    &                        & Be       \\
HIP 78207              & B8 & 0.0 & $-$5.6           & $-$21.37  &  abs    &  ipc      & ...   &  abs    &                        & Be       \\
HIP 77911              & B9 & 2.9 & $-$12.16         &  abs      &  abs    &  abs      & ...   &  abs    &                        &          \\
HIP 79410              & B9 & 0.6 & $-$9.4           &  abs      &  abs    &  abs      & ...   &  abs    &                        &          \\
HIP 79439              & B9 & 0.6 & $-$9.5           &  abs      &  abs    &  abs      & ...   &  abs    &                        &          \\
HIP 80024              & B9 & 0.7 & $-$9.25          &  abs      &  abs    &  abs      & 0.03:&  abs    &                        &          \\
HIP 76310              & A0 & 0.3 & $-$3.72          &  abs      &  abs    &  abs      & ...   &  abs    &                        &          \\
HIP 79156              & A0 & 0.5 & $-$9.17          &  abs      &  abs    &  abs      & ...   &  abs    &                        &          \\
HIP 79878              & A0 & 0.0 & $-$5.64          &  abs      &  abs    &  abs      & ...   &  abs    &                        &          \\
HIP 78996              & A9 & 0.3 & $-$12.52         &  abs      &  abs    &  abs      & ...   &  abs    &                        &          \\
HIP 80088              & A9 & 0.3 & $-$11.05         &  abs      &  abs    &  abs      & ...   &  abs    &                        &          \\
$[$PZ99$]$J161411.0-230536 & K0 & 2.4 &  ...         &  +0.80    &  +0.07  &  +0.38    & 0.41  &  $-$0.29 &                          & WTTS     \\
$[$PZ99$]$J160421.7-213028 & K2 & 1.0 & $-$6.27$\pm$1.46 &  cer      &  cer    &  $-$0.57  & 0.45 & $-$0.43 & 6300                     & WTTS       \\
$[$PZ99$]$J160357.6-203105 & K5 & 0.9 & $-$7.74$\pm$1.46 & $-$0.94   & $-$0.06 &  $-$11.57 & 0.49 & $-$0.57 & 6300                     & CTTS, accretor \\
J160643.8-190805       & K6 & 1.9 & $-$6.27$\pm$1.04 & $-$0.67   & $-$0.07 &  $-$2.39  & 0.51 & $-$0.59 &                        & WTTS       \\ 
J160900.7-190852       & K9 & 0.8 & $-$7.81$\pm$0.38 & $-$6.81   & $-$0.50 &  $-$20.08 & 0.50 & $-$0.75 & 6300, 6678             & CTTS, accretor \\
J160823.2-193001       & K9 & 1.5 & $+$5.83$\pm$2.99 & $-$2.28   & $-$0.18 &  $-$2.75  & 0.56 & $-$0.34 &                        & WTTS         \\
J161420.2-190648       & M0 & 1.8 & $-$7.25$\pm$2.16 & $-$7.25   & $-$1.07 &  $-$43.72 & 0.39 & $-$0.54 & 6300, 6678, 6717, 6731 & CTTS, accretor \\
ScoPMS 31              & M0.5 & 0.6 & $-$5.41$\pm$0.98 & $-$5.61 & $-$0.50 & $-$21.06  & 0.53 & $-$0.79 & 6300                     & CTTS, accretor \\
$[$PZ99$]$J155734.4-232111 & M1 & 0.8 & $-$4.63$\pm$2.00 & $-$2.41   & $-$0.16 &  $-$3.65  & 0.61 & $-$0.32 &                          & WTTS     \\
J161115.3-175721       & M1 & 1.6 & $-$7.77$\pm$0.90 & $-$2.19   & $-$0.18 &  $-$4.47  & 0.54 & $-$0.81 & 6300                   & WTTS     \\
J160545.4-202308       & M2 & 1.4 &     ...          & $+$1.04   & $-$0.13 &  $-$2.04  & 0.32 & $-$0.17 &                        & WTTS, SB2      \\
J160357.9-194210       & M2 & 1.7 & $-$3.82$\pm$0.83 & $-$2.43   & $-$0.20 &  $-$2.70  & 0.41 & $-$0.45 & 6300                   & WTTS     \\
J160953.6-175447       & M3 & 4.1 & $-$6.30$\pm$1.05 & $-$13.68  & $-$2.19 & $-$22.23  & 0.59 & $-$0.23 & 6300, 6678             & CTTS     \\
J155829.8-231007       & M3 & 0.0 & $-$5.63$\pm$0.89 & $-$55.00  & $-$8.83 & $-$158.48 & 0.31 & $-$0.84 & 6300, 6678             & CTTS, accretor \\
J155706.4-220606       & M4 & 2.0 & $-$6.11$\pm$1.74 & $-$8.43   & $-$0.49 &  $-$9.92  & 0.54 & $+$0.43 & 6300, 6678             & WTTS, accretor \\
J155729.9-225843       & M4 & 1.4 & $-$1.50$\pm$1.06 & $-$5.63   & $-$0.65 &  $-$6.91  & 0.59 & $-$0.05 & 6300, 6678             & WTTS         \\
J160959.4-180009       & M4 & 0.2 & $-$7.50$\pm$1.39 & $-$3.64   & $-$0.25 & $-$4.41   & 0.58 & $+$0.46 & 6300                   & WTTS         \\
J155624.8-222555       & M4 & 1.7 & $-$6.88$\pm$0.83 & $-$5.48   & $-$0.36 & $-$5.51   & 0.60 & $-$0.07 & 6300                   & WTTS         \\
J160525.5-203539       & M5 & 1.5 & $-$4.17$\pm$1.63 & $-$9.53   & $-$0.63 &  $-$8.61  & 0.62 & $+$0.42 & 6300, 6678             & WTTS         \\
J160532.1-193315       & M5 & 0.0 & $-$3.36$\pm$1.58 & $-$101.91 & $-$7.78 & $-$152.12 & 0.39 & $-$0.75 & 6300, 6678             & CTTS, accretor \\
J160611.9-193532       & M5 & 0.9 & $-$7.38$\pm$5.84 & $-$13.31  & $-$1.40 & $-$15.24  & 0.67 & $-$0.06 &                        & WTTS, multiple \\
J160622.8-201124       & M5 & 0.0 & $-$5.54$\pm$0.88 & $-$3.25   & $-$0.17 & $-$3.12   & 0.59 & $+$0.78 & 6300                   & WTTS         \\
J160702.1-201938       & M5 & 1.7 & $-$5.00$\pm$1.78 & $-$5.27   & $-$0.84 & $-$8.30   & 0.58 & $+$0.21 & 6678                   & WTTS         \\
J160827.5-194904       & M5 & 1.4 & $-$7.14$\pm$1.68 & $-$18.05  & $-$3.00 & $-$14.56  & 0.52 & $+$0.11 & 6678                   & WTTS         \\
J160900.0-190836       & M5 & 0.7 & $-$8.60$\pm$1.37 & $-$12.85  & $-$1.15 & $-$12.88  & 0.56 & $-$0.12 & 6678                   & WTTS         \\
J160600.6-195711       & M5 & 1.7 & $-$5.51$\pm$1.90 & $-$4.74   & $-$0.30 & $-$4.11   & 0.57 & $+$0.69 & 6678                   & WTTS         \\
\enddata
\tablenotetext{a}{Spectral type from the literature.}
\tablenotetext{b}{Extinction estimates are taken from Preibisch \& Zinnecker (1999) and Preibisch et al. (2002).}
\tablenotetext{c}{Radial velocity estimates from the literature for early-type (B+A) members and from cross-correlation analysis of HIRES observations for late-type (K+M) members.}
\tablenotetext{d}{Negative values indicate emission. abs - absorption, cer - core emission reversal, ipc - inverse P Cygni.}
\tablenotetext{e}{Negative values indicate emission. If core emission reversal is present, equivalent widths are measured from the base of the Ca II absorption profile.}
\tablenotetext{f}{[O I] $\lambda$6300, He I $\lambda$6678, [S II] $\lambda\lambda$6617, 6731.}
\tablenotetext{g}{Be: classical Be star; CTTS/WTTS determination using the spectral type dependent W(H$\alpha$) criteria of White \& Basri (2003); Accretion based upon H$\alpha$ velocity width analysis.}
\end{deluxetable}

\begin{deluxetable}{ccccc}
\tabletypesize{\tiny}
\tablenum{2}
\tablewidth{0pt}
\tablecaption{Median Spectral Energy Distributions}
\tablehead{
B9--A9 \\
\colhead{Wavelength}  &  \colhead{Median\tablenotemark{a}}      & \colhead{Lower Quartile\tablenotemark{a}} & \colhead{Upper Quartile\tablenotemark{a}} \\
        ($\mu$m)     &  log($\nu$F$_{\nu}$)   & log($\nu$F$_{\nu}$)      & log($\nu$F$_{\nu}$)      \\
}
\startdata
1.25     &  $-$8.28  &  $-$8.30  &  $-$8.27 \\
1.65     &  $-$8.60  &  $-$8.60  &  $-$8.60 \\
2.17     &  $-$8.88  &  $-$8.89  &  $-$8.87 \\
5.70     &  $-$10.04 &  $-$10.10 &  $-$10.03 \\
7.10     &  $-$10.34 &  $-$10.39 &  $-$10.32 \\
8.00     &  $-$10.56 &  $-$10.58 &  $-$10.52 \\
9.20     &  $-$10.67 &  $-$10.70 &  $-$10.62 \\
9.80     &  $-$10.71 &  $-$10.74 &  $-$10.65 \\
11.30    &  $-$10.86 &  $-$10.89 &  $-$10.81 \\
12.30    &  $-$10.90 &  $-$11.00 &  $-$10.89 \\
13.25    &  $-$10.97 &  $-$11.09 &  $-$10.94 \\
16.25    &  $-$11.25 &  $-$11.32 &  $-$11.16 \\
18.00    &  $-$11.24 &  $-$11.30 &  $-$11.18 \\
21.00    &  $-$11.28 &  $-$11.39 &  $-$11.25 \\
25.00    &  $-$11.34 &  $-$11.49 &  $-$11.29 \\
30.00    &  $-$11.46 &  $-$11.61 &  $-$11.27 \\
34.00    &  $-$11.53 &  $-$11.62 &  $-$11.16 \\
\\
\hline
\hline
K5--M2 \\
\colhead{Wavelength}  &  \colhead{Median\tablenotemark{a}}      & \colhead{Lower Quartile\tablenotemark{a}} & \colhead{Upper Quartile\tablenotemark{a}} \\
        ($\mu$m)     &  log($\nu$F$_{\nu}$)   & log($\nu$F$_{\nu}$)      & log($\nu$F$_{\nu}$)      \\
\hline
1.25     &   $-$9.09   &  $-$9.13   &  $-$9.06 \\
1.65     &   $-$9.16   &  $-$9.16   &  $-$9.16 \\
2.17     &   $-$9.36   &  $-$9.39   &  $-$9.32 \\
5.70     &   $-$10.18  &  $-$10.29  &  $-$10.12 \\
7.10     &   $-$10.40  &  $-$10.48  &  $-$10.33 \\
8.00     &   $-$10.49  &  $-$10.55  &  $-$10.41 \\
9.20     &   $-$10.37  &  $-$10.42  &  $-$10.10 \\
9.80     &   $-$10.33  &  $-$10.41  &  $-$10.04 \\
11.30    &   $-$10.35  &  $-$10.52  &  $-$10.11 \\
12.30    &   $-$10.50  &  $-$10.68  &  $-$10.30 \\
13.25    &   $-$10.55  &  $-$10.65  &  $-$10.33 \\
16.25    &   $-$10.48  &  $-$10.70  &  $-$10.23 \\
18.00    &   $-$10.30  &  $-$10.64  &  $-$10.10 \\
21.00    &   $-$10.27  &  $-$10.68  &  $-$10.11 \\
25.00    &   $-$10.29  &  $-$10.83  &  $-$10.18 \\
30.00    &   $-$10.41  &  $-$11.04  &  $-$10.33 \\
34.00    &   $-$10.44  &  $-$11.05  &  $-$10.32 \\
\\
\hline
\hline
M3--M5 \\
\colhead{Wavelength}  &  \colhead{Median\tablenotemark{a}}      & \colhead{Lower Quartile\tablenotemark{a}} & \colhead{Upper Quartile\tablenotemark{a}} \\
        ($\mu$m)     &  log($\nu$F$_{\nu}$)   & log($\nu$F$_{\nu}$)      & log($\nu$F$_{\nu}$)      \\
\hline
1.25     &   $-$10.41  &  $-$10.43  &  $-$10.35 \\
1.65     &   $-$10.49  &  $-$10.49  &  $-$10.49 \\
2.17     &   $-$10.65  &  $-$10.67  &  $-$10.64 \\
5.70     &   $-$11.55  &  $-$11.62  &  $-$11.40 \\
7.10     &   $-$11.77  &  $-$11.88  &  $-$11.58 \\
8.00     &   $-$11.78  &  $-$11.98  &  $-$11.67 \\
9.20     &   $-$11.72  &  $-$11.92  &  $-$11.51 \\
9.80     &   $-$11.70  &  $-$11.95  &  $-$11.52 \\
11.30    &   $-$11.77  &  $-$12.07  &  $-$11.68 \\
12.30    &   $-$11.84  &  $-$12.21  &  $-$11.70 \\
13.25    &   $-$11.85  &  $-$12.25  &  $-$11.74 \\
16.25    &   $-$12.04  &  $-$12.32  &  $-$11.85 \\
18.00    &   $-$11.90  &  $-$12.24  &  $-$11.69 \\
21.00    &   $-$11.94  &  $-$12.28  &  $-$11.67 \\
25.00    &   $-$12.00  &  $-$12.36  &  $-$11.66 \\
30.00    &   $-$12.04  &  $-$12.42  &  $-$11.67 \\
34.00    &   $-$11.99  &  $-$12.46  &  $-$11.62 \\
\enddata
\tablenotetext{a}{Units of median, lower, and upper quartiles are ergs s$^{-1}$ cm$^{-2}$}
\end{deluxetable}

\begin{deluxetable}{cccccc}
\tabletypesize{\tiny}
\tablenum{3}
\tablewidth{0pt}
\tablecaption{Stellar Properties and Mass Accretion Rates}
\tablehead{
\colhead{Source}  &  \colhead{$r$ (veiling)\tablenotemark{a}}  & \colhead{Mass\tablenotemark{b}}  & \colhead{Radius\tablenotemark{c}} &  \colhead{log $\dot{M}$\tablenotemark{d}} &  \colhead{log $\dot{M}$\tablenotemark{e}} \\
                                       &                 & (M$_{\odot}$)                  &           (R$_{\odot}$)             & log (M$_{\odot}$/yr)   & log (M$_{\odot}$/yr)    \\
}
\startdata
$[$PZ99$]$J160357.6-203105          &   0.16 &    1.07                       &             1.51                  &         $-$9.12                       &          $-$9.10                            \\
J160900.7-190852              &   0.1  &    0.58                       &             1.12                  &         $-$9.95                       &          $-$9.30                            \\
J161420.2-190648              &   1.0  &    0.58                       &             1.07                  &         $-$8.91                       &          $-$8.89                            \\
ScoPMS 31          &   0.05 &    0.50                       &             1.49                  &         $-$10.0                       &          $-$8.85                            \\
J155829.8-231007              &   1.50 &    0.27                       &             0.41                  &         $-$9.91                      &          $-$10.49                            \\
J155706.4-220606              &   0.39 &    0.30                       &             1.39                  &         $-$9.14                       &          $-$9.69                            \\
J160532.1-193315              &   0.18 &    0.11                       &             0.46                  &         $-$10.97                      &          $-$10.54                           \\
\enddata
\tablenotetext{a}{Defined such that $r = F_{exc}/F_{phot}$.}
\tablenotetext{b}{Stellar Mass (M$_{\odot}$) from the models of Siess et al. (2000).}
\tablenotetext{c}{Stellar Radius (R$_{\odot}$) from the models of Siess et al. (2000).}
\tablenotetext{d}{Mass accretion rate derived from the optical veiling analysis near $\lambda$6500.}
\tablenotetext{e}{Mass accretion rate derived from the log-linear relation between Ca II $\lambda$8542 emission line luminosity and accretion luminosity from Herczeg \& Hillenbrand (2008).}
\end{deluxetable}

\begin{deluxetable}{ccccc}
\tabletypesize{\tiny}
\tablenum{4}
\tablewidth{0pt}
\tablecaption{Mid-Infrared Spectral Indices}
\tablehead{
\colhead{Source}  &  \colhead{n$_{6-13}$\tablenotemark{a}}  & \colhead{n$_{13-25}$\tablenotemark{a}} & \colhead{n$_{6-25}$\tablenotemark{a}} & \colhead{$(F_{10}-F_{cont})/F_{cont}$\tablenotemark{b}} \\
}
\startdata
HIP 80024 & $-$2.66 & $-$0.76 & $-$1.84 & 0.01 \\
HIP 80088 & $-$2.73 & $-$1.09 & $-$2.02 & 0.02 \\
HIP 79878 & $-$2.57 & $-$1.65 & $-$2.17 & $-$0.07 \\
HIP 79439 & $-$2.76 & $-$2.27 & $-$2.55 & $-$0.05 \\
HIP 79410 & $-$2.33 & $-$1.56 & $-$2.00 & $-$0.06 \\
HIP 79156 & $-$2.53 & $-$2.03 & $-$2.31 & $-$0.02 \\
HIP 78996 & $-$2.50 & $-$1.50 & $-$2.07 & $-$0.05 \\
HIP 76310 & $-$2.17 & $+$0.83 & $-$0.89 & $-$0.04 \\
$[$PZ99$]$J161411.0-230536 & $-$0.44 & $-$1.39 & $-$0.85 & 0.16 \\
$[$PZ99$]$J160421.7-213028 & $-$2.69 & $-$0.76 & $-$1.84 & 0.12 \\
$[$PZ99$]$J160357.6-203105 & $-$0.78 & $+$0.43 & $-$0.26 & $+$0.31 \\
J161420.2-190648     & $-$0.58 & $-$0.07 & $-$0.36 & $+$0.84 \\
J161115.3-175721     & $-$0.97 & $-$1.08 & $-$1.02 & $+$0.63 \\
J160900.7-190852     & $-$0.09 & $+$0.40 & $+$0.12 & $+$0.78 \\
J160823.2-193001     & $-$0.65 & $-$0.35 & $-$0.52 & $+$0.27 \\
J160702.1-201938     & $-$1.51 & $-$0.67 & $-$1.15 & $+$0.30 \\
J160643.8-190805     & $-$2.15 & $-$1.45 & $-$1.85 & $+$0.01 \\
J160357.9-194210     & $-$1.41 & $-$0.40 & $-$0.98 & $+$0.66 \\
ScoPMS 31            & $-$0.87 & $+$1.31 & $+$0.07 & $+$0.36 \\
J160959.4-180009     & $-$0.36 & $+$1.17 & $+$0.30 & $+$1.08 \\
J160953.6-175446     & $-$1.15 & $-$0.69 & $-$0.95 & $+$0.22 \\
J160900.0-190836     & $-$0.85 & $+$0.39 & $-$0.32 & $+$0.53 \\
J160827.5-194904     & $-$1.58 & $-$1.12 & $-$1.38 & $+$0.35 \\
J160622.8-201124     & $-$1.02 & $+$0.54 & $-$0.36 & $+$0.70 \\
J160600.6-195711     & $-$1.89 & $-$0.26 & $-$1.18 & $+$0.30 \\
J160532.1-193315     & $-$0.72 & $-$2.90 & $-$1.66 & ...     \\
\enddata
\tablenotetext{a}{Spectral indices defined in \S 5.3, from 6 to 13 $\mu$m, 13 to 25 $\mu$m, and 6 to 25 $\mu$m}
\tablenotetext{b}{The continuum subtracted, integrated flux of the 10 $\mu$m silicate emission feature normalized to the continuum.} 
\end{deluxetable}

\begin{deluxetable}{cccccc}
\tabletypesize{\tiny}
\tablenum{5}
\tablewidth{0pt}
\tablecaption{Dust Properties of Early-Type Excess Stars Inferred from Photosphere-Subtracted Spectral Energy Distributions}
\tablehead{
\colhead{Source}  & \colhead{$T_{gr}$\tablenotemark{a}} & \colhead{$L_{IR}/L_{*}$\tablenotemark{b}} & \colhead{$r_{min}$\tablenotemark{c}} & \colhead{$a_{rad}$\tablenotemark{d}} & \colhead{$M_{dust}$\tablenotemark{e}} \\
                  &    (K)                    &                                & (AU)               & ($\mu$m)                     & (M$_{\earth}$) \\ 
}
\startdata
HIP 76310 &    122  & 2.1$\times$10$^{-4}$ & 29  & 3.0  & 8.2$\times$10$^{-5}$ \\
HIP 78996 &    150  & 7.7$\times$10$^{-5}$ & 11  & 1.4  & 2.0$\times$10$^{-6}$ \\
HIP 79156 &    150  & 7.7$\times$10$^{-5}$ & 12  & 1.4  & 1.1$\times$10$^{-6}$ \\
HIP 79410 &    195  & 2.6$\times$10$^{-5}$ & 15  & 4.3  & 4.0$\times$10$^{-6}$ \\
HIP 79439 &    150  & 5.6$\times$10$^{-6}$ & 26  & 4.8  & 2.9$\times$10$^{-6}$ \\
HIP 79878 &    170  & 2.6$\times$10$^{-5}$ & 15  & 3.0  & 2.8$\times$10$^{-6}$ \\
HIP 80024 &    105  & 5.8$\times$10$^{-5}$ & 55  & 3.7  & 1.0$\times$10$^{-4}$ \\
HIP 80088 &    120  & 8.8$\times$10$^{-5}$ & 13  & 1.0  & 2.4$\times$10$^{-6}$ \\
\enddata
\tablenotetext{a}{The blackbody temperature used to fit the photosphere-subtracted IRS spectrum.}
\tablenotetext{b}{The fractional infrared luminosity of the source.}
\tablenotetext{c}{The minimum orbital radius of the dust grains.}
\tablenotetext{d}{The dust grain blow-out radius due to stellar radiation pressure.}
\tablenotetext{e}{The minimum dust mass responsible for the observed excess emission.}

\end{deluxetable}

\end{document}